\documentclass[11pt,a4paper]{article}
\usepackage{facsat}

\newcommand{\Fesc}{139}

\newcommand{\pymsisver}{0.12.0}
\newcommand{\pythonver}{3.14}

\begin{document}

\thispagestyle{plain}

\begin{center}
{\LARGE\bfseries Premature re-entry of the FACSAT-2 nanosatellite:\\[2pt]
orbital decay and attribution to the maximum of solar cycle 25\par}

\vspace{1.2em}

{\large Santiago Vargas Dom\'inguez\,$^{\ast}$, \quad Mara Valentina Angel, \quad Brayan Nicolas Buitrago\par}

\vspace{0.9em}

{\small
Observatorio Astron\'omico Nacional, Universidad Nacional de Colombia, Bogot\'a, Colombia\\[3pt]
$^{\ast}$\,Corresponding author: \texttt{svargasd@unal.edu.co}\\[3pt]
}
\end{center}

\vspace{0.6em}
\vspace{1.0em}

\begin{resumenbox}{Abstract}
FACSAT-2 \emph{Chiribiquete}, the second satellite of the Colombian Aerospace Force and the first
South American nanosatellite mission dedicated to greenhouse gases, re-entered on 27 October 2025,
926 days after its deployment into a Sun-synchronous orbit, against a declared service life of at
least five years and an estimated orbital lifetime of fifteen. The orbital evolution was built from
the \texttt{epochs} of the TLEs (two-line elements) of NORAD ID 56205. We reconstruct its decay using
its 3696 orbital elements, the solar and geomagnetic indices from the OMNI2 database, and
NRLMSISE-00 densities sampled along the orbit. Inverting the drag yields a ballistic coefficient of
\SI{72.3}{\kilo\gram\per\meter\squared} (\SI{6}{\percent} statistical, \SI{20}{\percent}
systematic), equivalent to an effective area of \SI{0.047}{\meter\squared}, consistent with the
declared geometry. With it, the model reproduces the trajectory with a root-mean-square error of
\SI{7.4}{\kilo\meter} and, calibrated on 2023--2024 alone, the re-entry with a 28-day out-of-sample
error. \Fdiez{} averaged \SI{170.8}{\sfu} over the mission, against the \SI{\Fesc}{\sfu} of the
environmental scenario derived from the official 2019 forecast, which lengthens the orbital lifetime
by $\times1.52$; two co-launched CubeSats give $\times1.50$ and $\times1.57$, a consistency check
across objects rather than independent evidence. Imposing quiet geomagnetic conditions returns 102
days, \SI{22}{\percent} of the shortening, with strong sensitivity to the adopted threshold; a
superposed-epoch analysis over 149 storms gives a median drag enhancement of $\times1.92$ for the
intense events. We discuss the consequences for the design of FACSAT-3 and for the public
communication of a mission's expected lifetime.

\vspace{4pt}\noindent\textbf{Keywords:} FACSAT-2, orbital decay, atmospheric drag, solar cycle 25,
\mbox{space weather}, CubeSat.
\end{resumenbox}

\vspace{0.6em}

%%%%%%%%%%%%%%%%%%%%%%%%%%%%%%%%%%%%%%%%%%%%%%%%%%%%%%
\section{Introduction}
\label{sec:intro}

FACSAT-2 \emph{Chiribiquete} was the second satellite of the Colombian Aerospace Force and the first
South American nanosatellite mission dedicated to the analysis of greenhouse gases
\citep{rincon2025}. It was a 6U CubeSat of \SI{7.49}{\kilo\gram} \citep{fac_factibilidad} equipped
with a multispectral camera of \SI{4.75}{\meter} resolution and \SI{19.4}{\kilo\meter} swath
\citep{simera2021}, deployed into a Sun-synchronous orbit on 15 April 2023 from the Transporter-7
mission. It re-entered the atmosphere on 27 October 2025, 926 days later, against a declared service
life of at least five years and an estimated orbital lifetime of fifteen \citep{fac_facsat2}.

The discrepancy is not an isolated case. Solar cycle 25 comfortably exceeded the maximum of 115
smoothed sunspots that the NOAA/NASA/ISES panel had forecast in 2019 \citep{swpc_cycle2019}, and
extreme ultraviolet radiation controls the temperature and therefore the density of the thermosphere
\citep{emmert2015}. A cycle more active than expected consequently shortens the orbital lifetime of
any unpropelled object in low Earth orbit. A particularly relevant context is provided by CIRBE and
TAIFA-1, two 3U CubeSats deployed on the same day and into the same orbit, which re-entered on 3
October 2024 and 23 April 2025 respectively.

This work reconstructs the decay of FACSAT-2 from the complete archive of its orbital elements and
confronts it with the measured solar and geomagnetic indices, with four aims: to determine whether
the observed trajectory admits a description consistent with a nominal vehicle subjected to the
cycle that actually occurred, or whether it requires postulating some anomaly; to measure how much
orbital life was lost relative to what could have been expected at the design stage; to separate,
using measured indices, the contribution of sustained ultraviolet heating from that of discrete
storms, a distinction that public discussion tends to overlook; and to derive verifiable design
margins for FACSAT-3.

%%%%%%%%%%%%%%%%%%%%%%%%%%%%%%%%%%%%%%%%%%%%%%%%%%%%%%
\section{Data}
\label{sec:datos}

For FACSAT-2 ``Chiribiquete'' we used the object identified with NORAD ID 56205. Its orbital
evolution was reconstructed from its TLE history. The empirical basis is the complete archive of
two-line orbital elements of object 56205 published by the 18\textsuperscript{th} Space Defense
Squadron \citep{spacetrack2025}, made up of 3696 sets between 18 April 2023 and 27 October 2025 with
a median cadence of \SI{6.2}{\hour}. From each set we use the mean motion, the eccentricity, the
inclination and the $B^{*}$ coefficient, and from the first of these the semi-major axis is obtained
through $a = (\mu/n^{2})^{1/3}$, with the caveat that these are mean elements in SGP4 theory and not
osculating ones \citep{vallado2006}. To these we add the daily and hourly series of \Fdiez{}, $a_p$,
$K_p$ \citep{matzka2021} and Dst from the OMNI2 database \citep{king2005}, a storm catalogue built
from them, and a sampling of atmospheric density obtained with NRLMSISE-00 \citep{picone2002} along
the orbit. Table~\ref{tab:params} summarises the derived quantities.

The problem admits two altitude conventions that differ by more than ten kilometres. The mean
spherical altitude, $a$ minus the Earth's equatorial radius, is \SI{497.0}{\kilo\meter} at the start
of the mission; the orbit-averaged geodetic altitude, which is the one that determines the density
encountered, is \SI{509.1}{\kilo\meter}. The \SI{500}{\kilo\meter} of the payload specification
\citep{simera2021,rincon2025} and the \SI{508}{\kilo\meter} of the mission documentation are
consistent with this second definition and, in any case, with a nominal injection close to
\SI{500}{\kilo\meter}. Each figure states which one it uses. Section~S1 details this convention and
the temporal one, which distinguishes the \num{922.7} days of the archive, the \num{926.1} days
since deployment and the simulated orbital lifetime.

The storm catalogue was built with a simple and reproducible criterion: over the hourly Dst series
we identified the contiguous runs with $\mathrm{Dst} \le \SI{-50}{\nano\tesla}$, each run counting
as one event and its date being the hour of the minimum, without imposing a minimum separation or
merging multiple minima. The procedure returns 149 events: six with
$\mathrm{Dst} < \SI{-150}{\nano\tesla}$, twenty-eight between $-150$ and \SI{-80}{\nano\tesla} and
one hundred and fifteen above that. Since no minimum separation is imposed, the windows of the
superposed-epoch analysis overlap between nearby events, which motivates the robustness test of
Section~\ref{sec:tormentas}.

The comparison with other objects rests on CIRBE and TAIFA-1, deployed into the same orbit and on
the same day as FACSAT-2, so that the differences between their orbital lifetimes should be
dominated by their ballistic properties. FACSAT-1 serves as an intergenerational control: its
insertion elements are published \citep{portilla2021} and correspond to \SI{497.8}{\kilo\meter}
geodetic; it re-entered on 3 June 2023 after 4.51 years of flight begun in the depths of solar
minimum, six months short of the projected service life.

\begin{table}[htbp]
\centering
\caption{Quantities derived from the FACSAT-2 archive of orbital elements (NORAD 56205, COSPAR
2023-054AD). The revolution counter does not start at deployment: the number of revolutions is
obtained by difference and verified by integrating the mean motion.}
\label{tab:params}
\small
\begin{tabular}{@{}p{0.40\textwidth}p{0.52\textwidth}@{}}
\toprule
\textbf{Quantity} & \textbf{Value} \\
\midrule
Available elements & 3696, from 18 Apr 2023 to 27 Oct 2025, median cadence \SI{6.2}{\hour} \\
Median inclination & \ang{97.365} \\
First element & $n = 15.2292$ rev/day, $P = \SI{94.55}{\minute}$, $e = 0.00146$ \\
Initial altitude (mean spherical) & \SI{497.0}{\kilo\meter}, perigee \SI{487.3}{\kilo\meter}, apogee \SI{506.8}{\kilo\meter} \\
Initial altitude (orbital geodetic) & \SI{509.1}{\kilo\meter} \\
Last element & $n = 16.3959$ rev/day, $P = \SI{87.83}{\minute}$, $e = 0.00057$ \\
Revolutions in the archive & \num{14241} by difference, \num{14250} by integrating $n$ \\
Interval covered by the archive & \num{922.7} days (2.526 years) \\
Lifetime since deployment & \num{926.1} days (2.536 years), from 15 Apr 2023 06:47 UTC \\
Mission-mean \Fdiez{} & \SI{170.8}{\sfu}, monthly maximum \SI{253}{\sfu} in August 2024 \\
Mean $a_p$, minimum Dst & \SI{12.4}{\nano\tesla} and \SI{-406}{\nano\tesla} on 11 May 2024 \\
\bottomrule
\end{tabular}
\end{table}

%%%%%%%%%%%%%%%%%%%%%%%%%%%%%%%%%%%%%%%%%%%%%%%%%%%%%%
\section{Methodology}
\label{sec:metodo}

The methodology combines orbital, atmospheric and solar--geomagnetic activity information in order
to reconstruct the decay of FACSAT-2 and to characterise physically the drag experienced during the
mission. To that end we used the TLEs together with SGP4 propagation, the NRLMSISE-00 atmospheric
model and the solar and geomagnetic activity indices. From these sources the orbital evolution was
reconstructed and an effective ballistic coefficient was estimated independently of any direct
interpretation of \(BSTAR\).

The elements were cleaned by removing duplicates by epoch, resampled to daily resolution through the
median of those available on each day, gaps without elements were linearly interpolated up to a
maximum duration of three days, which is the longest run present in the archive, and the altitude
was smoothed with a centred seven-day moving window. The decay rate is the discrete daily derivative
of that smoothed series.

For a nearly circular orbit, a condition that FACSAT-2 amply satisfied, the drag equation
\citep{kinghele1987}
\begin{equation}
\frac{\mathrm{d}a}{\mathrm{d}t} = -\rho\,\frac{C_D A}{m}\,\sqrt{\mu\,a}
\label{eq:decay}
\end{equation}
allows the effective ballistic coefficient to be defined and solved for from the observation,
\begin{equation}
B_{\mathrm{eff}} \equiv \frac{m}{C_D A} = \frac{\rho\,\sqrt{\mu\,a}}{-\,\mathrm{d}a/\mathrm{d}t}.
\label{eq:beff}
\end{equation}
The estimate therefore rests on the evolution of the semi-major axis, its derivative and the density
actually encountered. $B^{*}$ was not used for it; although it carries information about the drag,
it is a parameter fitted within the SGP4 propagator and not a physical measurement of
$m/(C_D A)$, as Section~S5 and Figure~1S of the supplementary material document.

The smoothing window is not innocuous, because the drag response to a storm lasts little more than a
day. With the daily derivative over the seven-day smoothed series, the enhancement of 11 May 2024
reaches $\times3.38$ and the intense storms give a median of $\times1.92$; wider windows average the
phenomenon away entirely, as Section~S3 quantifies. The amplitudes of
Section~\ref{sec:tormentas} are only comparable with those of works that use an equivalent
smoothing.

\subsection{Atmospheric density}
\label{sec:densidad}

The thermospheric density was computed with NRLMSISE-00 \citep{picone2002} through the Python
package \texttt{pymsis} \citep{pymsis2022}, version \pymsisver{}, under Python~\pythonver{}. The
model version is set explicitly to NRLMSISE-00 (option \texttt{version=0}) and is not left at the
library default, which corresponds to NRLMSIS~2.1. The temporal convention for the indices is the
model's standard one, with the previous day's \Fdiez{}, the centred 81-day mean and the full
seven-component $a_p$ vector detailed in Section~S2.

For each day the representative element chosen was the one occupying position $\lfloor n/2 \rfloor$
in the chronological sequence of that day, counting from zero; on days with an even number of
elements the later of the two central ones is taken. The orbit was propagated with SGP4 from its
epoch and eight points equally spaced in time were sampled along one revolution, from which geodetic
latitude, longitude and altitude on the WGS84 ellipsoid were obtained. The harmonic decomposition of
the density variation along a revolution is dominated by the first two modes, and the error of the
average stays below \SI{1}{\percent} in all the altitude bands traversed (Section~S2). The densities
were combined into a daily orbital mean density $\rho(t)$ over 910 days.

\subsection{Ballistic coefficient and simulations}
\label{sec:beff_metodo}

With the series $a(t)$, $\mathrm{d}a/\mathrm{d}t$ and $\rho(t)$ a daily series of
$B_{\mathrm{eff}}$ of 909 usable values was built. Rather than imposing a single value from the
outset, its variation in time and by altitude band was examined, and only afterwards was it
summarised by the median. Its statistical uncertainty was estimated by moving-block bootstrap with
blocks of fifteen, thirty and sixty days, which is necessary because $B_{\mathrm{eff}}(t)$ retains
temporal structure; the systematic uncertainty, much larger, comes from the absolute density level
and was propagated by Monte Carlo with the model of Section~S4. Finally, physical plausibility was
checked by transforming the coefficient into an effective area
$A_{\mathrm{eff}} = m/(C_D B_{\mathrm{eff}})$ and contrasting it with the geometry of a 6U CubeSat.

The orbital lifetime simulations use a Jacchia-71 type density model \citep{jacchia1971} with
exospheric temperature
\begin{equation}
T_\infty = 379 + 3.24\,\overline{F}_{10.7} + 1.3\,(F_{10.7}-\overline{F}_{10.7})
           + 28\,K_p + 0.03\,e^{K_p} \quad [\si{\kelvin}],
\label{eq:tinf}
\end{equation}
the thermal profile of \citet{bates1959} and diffusive equilibrium above \SI{125}{\kilo\meter}. That
model carries biases with respect to the established climatology \citep{picone2002,emmert2015},
corrected by means of a transfer function fitted against the NRLMSISE-00 sampling,
\begin{equation}
\ln\frac{\rho_{\mathrm{Jacchia}}}{\rho_{\mathrm{MSIS}}}
 = -0.523 - 0.203\,\frac{h-450}{100} + 0.132\,\frac{\overline{F}_{10.7}-170}{50}
   + 0.023\,\frac{a_p-12}{20},
\label{eq:correccion}
\end{equation}
with $h$ in kilometres. The simplified model is used rather than NRLMSISE-00 directly because the
counterfactual scenarios require tens of thousands of evaluations spanning decades. The price of
that choice is bounded by applying the residual of the correction,
$\sigma[\ln(\rho_{\mathrm{Jacchia}}/\rho_{\mathrm{MSIS}})] = 0.17$, as if it were a global bias:
shifting the density by $\pm\SI{17}{\percent}$ moves the simulated orbital lifetime between 2.09 and
3.35 years and the shortening factor between $\times1.41$ and $\times1.57$. That \SI{17}{\percent}
is not a coherent bias but the point-by-point logarithmic scatter of the residuals, of zero mean by
construction, so treating it as a constant bias is a conservative bound. The correction transfers
the sensitivities of NRLMSISE-00 to the model, equivalent to a factor of $3.15$ in density per
\SI{100}{\sfu} of flux and a scale height of \SI{59}{\kilo\meter}; Table~3S of the supplementary
material collects the remaining parameters.

Equation~\eqref{eq:decay} is integrated with a daily step and second-order Runge--Kutta from the
real initial conditions, fed day by day with the measured indices and with the ballistic coefficient
fixed at the measured value; the only remaining degree of freedom, a global density scale factor, is
adjusted to reproduce the re-entry date and turns out to be $1.06$. Section~S8 verifies, with a
density field fitted to those same orbital densities, that the solution barely depends on the
integration step or on the re-entry threshold. The counterfactual scenarios replace the measured
indices with other assumed ones without touching anything else. The scenario derived from the
official 2019 forecast is built in three steps. First, the smoothed sunspot number is modelled with
the normalised shape function of \citet{hathaway2015},
\begin{equation}
R(t) = A\,\frac{g(x)}{g(x_{\max})}, \qquad g(x) = \frac{x^{3}}{e^{x^{2}}-0.71},
\qquad x = \frac{t-t_0}{b},
\label{eq:hathaway}
\end{equation}
where $g$ reaches its maximum at $x_{\max} = 1.081$, so that $A$ is directly the maximum of the
sunspot number and $A = 115$ reproduces the value announced by the panel \citep{swpc_cycle2019}.
Placing that maximum in July 2025 and the start of the cycle in April 2020, both dates from the
forecast itself, fixes the width at $b = 4.86$ years, so that the equation contains no parameter
fitted to the observed activity. Second, the curve is converted to flux through the empirical
relation between sunspot number and \Fdiez{} \citep{hathaway2015,clette2021}. Third, the geomagnetic
activity is set with a climatological relation tied to the flux that reproduces the mean
$a_p \approx \SI{12}{\nano\tesla}$ of the maximum. Section~S2 collects both expressions and the
conversion table between $a_p$ and $K_p$ \citep{matzka2021}.

The curve built in this way admits two averages that should not be confused. Over the window
actually flown, from 15 April 2023 to 27 October 2025, the scenario returns \SI{\Fesc}{\sfu}, which
is the figure homologous to the \SI{170.8}{\sfu} measured and the only one that admits direct
comparison with them; over the full life of the counterfactual trajectory, 3.83 years, it returns
\SI{135}{\sfu}, a lower value because that interval incorporates more of the descending branch of
the cycle. Both are results of the procedure and not imposed parameters.

From the same archive the precession rate of the ascending node is derived, which
Section~\ref{sec:discusion} uses to describe the degradation of the observing capability, through
the first-order secular approximation in $J_2$,
\begin{equation}
\frac{\mathrm{d}\Omega}{\mathrm{d}t} = -\frac{3}{2}\,J_2
\left(\frac{R_E}{a\,(1-e^{2})}\right)^{\!2} n \cos i ,
\label{eq:nodal}
\end{equation}
evaluated day by day with the mean elements of each day. The drift of the local solar time of the
node is obtained by integrating the difference between that rate and the ideal Sun-synchronous value
of \ang{0.9856} per day.

\subsection{Statistical treatment and validation}
\label{sec:estadistica}

The daily series of drag and of indices are strongly autocorrelated: the 909 values are not
equivalent to 909 independent observations, and a conventional Pearson probability would
underestimate the null level. Significances are therefore evaluated through an effective sample size
derived from the first-order autocorrelations and through a block bootstrap with thirty-day blocks
in which the blocks of one series are resampled independently of those of the other, so that the
internal autocorrelation is preserved but the association between the two is destroyed under the
null hypothesis. The significance of the periodogram is estimated against random permutations and
against phase surrogates, a considerably more demanding test. Section~S4 details the procedure and
the randomisation options.

The dynamical reconstruction was additionally subjected to an out-of-sample test: the global density
factor was recalibrated with the orbital data up to 31 December 2024, with the satellite still at
\SI{429}{\kilo\meter}, and with that calibration the rest of the mission was propagated without
touching any parameter. The subsequent propagation is fed with the \Fdiez{} and $a_p$ actually
observed during 2025, so it is a \emph{hindcast} conditioned on the observed forcing and not a
prospective prediction; what is being tested is the dynamics and its calibration, not the
predictability of space weather.

To separate the slow forcing from the fast episodes, the drag anomaly is defined as the ratio
between the observed value of $\rho/B$ and a baseline obtained by a centred 81-day moving median. It
matters to stress what that signal is made of, because the validity of all the subsequent
correlations depends on it. The observational quantity is
\begin{equation}
\frac{\rho}{B} = -\frac{1}{\sqrt{\mu\,a}}\,\frac{\mathrm{d}a}{\mathrm{d}t},
\label{eq:rhoB}
\end{equation}
which is obtained solely from the semi-major axis and its time derivative. The anomaly used in the
correlations with \Fdiez{}, $a_p$, $K_p$ and Dst is therefore built exclusively from the observed
orbital decay, without using NRLMSISE-00 or any index: the atmospheric model intervenes only when
converting that ratio into a ballistic coefficient and in the simulations, so the correlations
cannot inherit a dependence on the indices by that route.

%%%%%%%%%%%%%%%%%%%%%%%%%%%%%%%%%%%%%%%%%%%%%%%%%%%%%%
\section{Results}
\label{sec:resultados}

\subsection{Trajectory and ballistic coefficient}
\label{sec:balistico}

\begin{figure}[htbp]
\centering
\includegraphics[width=0.98\textwidth]{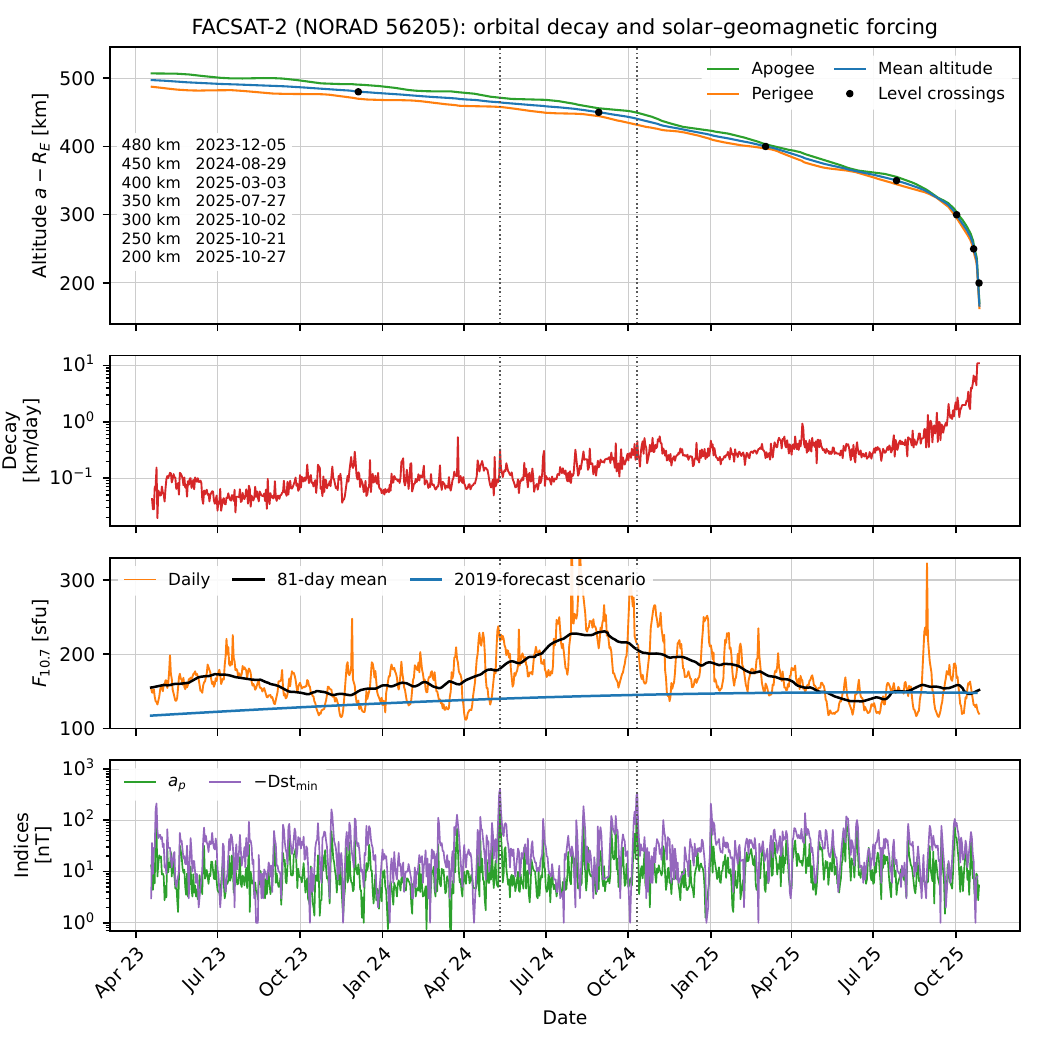}
\caption{Overview of the mission. From top to bottom, apogee, perigee and mean altitudes derived
from the 3696 orbital elements with the crossing dates of each level; daily decay rate on a
logarithmic scale; daily \Fdiez{} flux, its 81-day mean and the \Fdiez{} of the scenario derived
from the official 2019 forecast \citep{swpc_cycle2019}; and the $a_p$ and $-\mathrm{Dst}_{\min}$
indices \citep{king2005}. The vertical dotted lines mark the storms of May and October 2024.
Altitudes in the equivalent spherical convention $a-R_E$.}
\label{fig:panorama}
\end{figure}

Figure~\ref{fig:panorama} summarises the mission. The descent was slow during the first year, with
rates below \SI{0.1}{\kilo\meter\per\day}, and accelerates without interruption until it exceeds
\SI{10}{\kilo\meter\per\day} in the final days: the satellite took 234 days to cover the
\SI{100}{\kilo\meter} that separate \SI{400}{\kilo\meter} from \SI{300}{\kilo\meter} and only 25 to
cover the remaining \SI{180}{\kilo\meter} down to re-entry.

The direct inversion of Equation~\eqref{eq:beff} yields a median of
$B_{\mathrm{eff}} = \SI{72.3}{\kilo\gram\per\meter\squared}$ over 909 usable days.
Figure~\ref{fig:ballistic} shows that the value stays between \num{67.9} and
\SI{79.9}{\kilo\gram\per\meter\squared} when broken down by altitude band, without a systematic
trend and with overlapping intervals, which supports the hypothesis of an approximately constant
coefficient over \SI{250}{\kilo\meter} of descent.

The autocorrelation of the series decays with a scale of five days: the 909 values are equivalent to
some 190 independent observations, and an independent bootstrap would underestimate the interval.
With moving blocks the half-width goes from \SI{2.2}{\percent} to \SI{4.5}{\percent} with fifteen-day
blocks, to \SI{6.0}{\percent} with thirty and to \SI{7.8}{\percent} with sixty, without the median
moving; the \SI{6}{\percent} of the thirty-day blocks is adopted. The dominant uncertainty is
nevertheless systematic and comes from the absolute density level of NRLMSISE-00, whose typical
error is of the order of \SI{20}{\percent} \citep{emmert2015}; propagating it, the coefficient falls
between \num{58} and \SI{87}{\kilo\gram\per\meter\squared}.

\begin{figure}[htbp]
\centering
\includegraphics[width=0.94\textwidth]{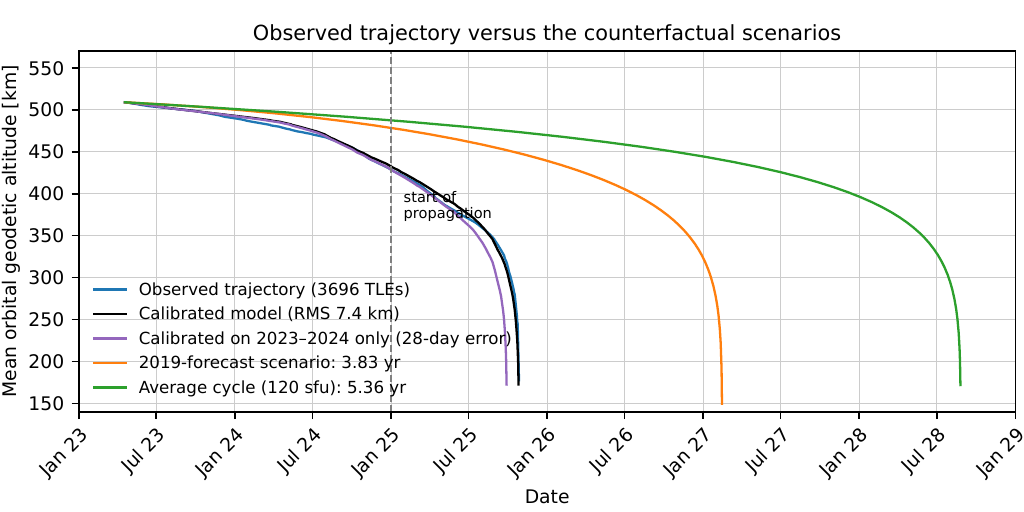}
\caption{Observed trajectory, model with the ballistic coefficient fixed at
$B_{\mathrm{eff}}=\SI{72.3}{\kilo\gram\per\meter\squared}$, and counterfactual scenarios derived
from the 2019 solar forecast and from an average cycle. Both scenarios modify \Fdiez{} and the
geomagnetic activity climatologically tied to it at the same time. Altitudes in the mean orbital
geodetic convention.}
\label{fig:decay}
\end{figure}

\begin{figure}[htbp]
\centering
\includegraphics[width=0.98\textwidth]{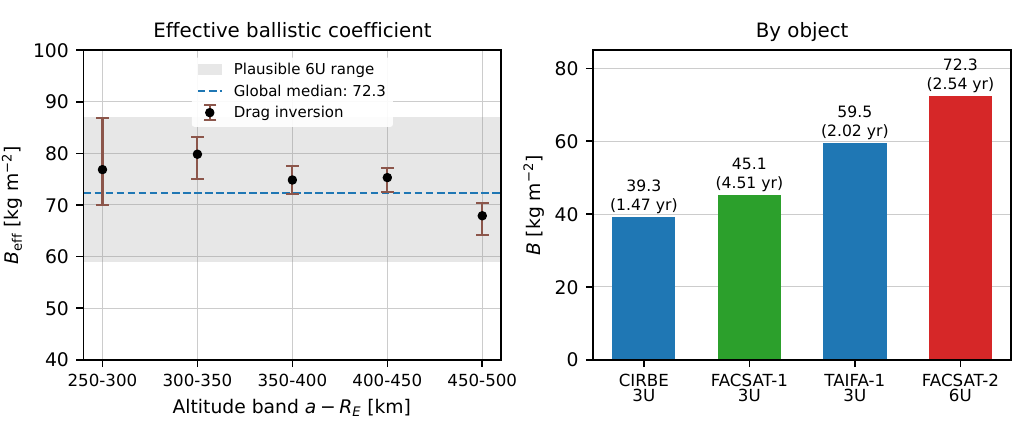}
\caption{Left, effective ballistic coefficient by altitude band, in the equivalent spherical
convention $a-R_E$, with \SI{95}{\percent} bootstrap intervals. The shaded band is the range of
plausible projected areas for this vehicle, not a strict geometric limit: it comprises the
coefficients compatible with areas between \num{0.043} and \SI{0.060}{\meter\squared} for the
nominal mass and $C_D = 2.2$, where the upper end is the largest face of the bus and the lower one
captures the effect of varying the drag coefficient up to $C_D = 2.4$. As a geometric reference, a
6U of $0.10 \times 0.20 \times \SI{0.30}{\meter}$ projects \num{0.020}, \num{0.030} and
\SI{0.060}{\meter\squared} depending on the face and \SI{0.055}{\meter\squared} in random tumbling.
The dashed line is the global median. Right, ballistic coefficient of the four objects analysed,
with the lifetime since deployment in parentheses.}
\label{fig:ballistic}
\end{figure}

Fixing that coefficient, the integration of Equation~\eqref{eq:decay} reproduces the \num{922.7}
days of trajectory with a root-mean-square error of \SI{7.4}{\kilo\meter} and a single degree of
freedom (Figure~\ref{fig:decay}). It matters to be precise about what that agreement demonstrates,
namely that the two branches of the calculation are not fully independent, because both draw on
NRLMSISE-00 and because the global factor is fitted to the re-entry date. What it establishes is
that the trajectory admits a description with a single physical parameter and a level correction of
\SI{6}{\percent}, without invoking any anomaly of the vehicle.

The out-of-sample test is more demanding. Recalibrating the density factor using only the data up to
the end of 2024, when the satellite had ten months of flight left, and propagating afterwards
without touching anything, the model places the re-entry on 29 September 2025 against the real 27
October: an error of 28 days, \SI{3}{\percent} of the orbital lifetime, with a root-mean-square
error of \SI{31}{\kilo\meter} in the phase not used for calibration against
\SI{2.8}{\kilo\meter} in the fitting stretch. The exercise is conditioned on the forcing observed in
2025, so that it validates the dynamics and its calibration but not the predictability of the space
environment. With that caveat, it supports the central argument: to explain the re-entry a nominal
6U exposed to the cycle that actually occurred is enough.

The physical plausibility check closes the argument. With the nominal mass of
\SI{7.49}{\kilo\gram} \citep{fac_factibilidad} and $C_D = 2.2$, the coefficient implies an effective
area of \SI{0.047}{\meter\squared}. Propagating by Monte Carlo the density bias, the drag
coefficient and the mass tolerance, the median holds with a \SI{68}{\percent} central interval
between \num{0.038} and \SI{0.058}{\meter\squared}, contained between the projected areas of the
smaller and the larger face of a 6U. \SI{77}{\percent} of the realisations fall below the
\SI{0.055}{\meter\squared} of random tumbling, a fraction that should not be read as a probability
in the strict sense because the density uncertainty that generates it does not constitute a
well-established distribution.

The four objects analysed order themselves by ballistic coefficient in agreement with their size and
mass, from the \SI{39.3}{\kilo\gram\per\meter\squared} of CIRBE to the
\SI{72.3}{\kilo\gram\per\meter\squared} of FACSAT-2, which constitutes an internal control of the
method (Table~1S of the supplementary material). The archive also allows the evolution of the other
elements to be followed, collected in Section~S6 and in Figure~2S: the inclination fell by
\ang{-0.128}, the eccentricity decays from $0.00146$ to $0.00057$ with a 99-day apsidal cycle
consistent with the rotation of the line of apsides, and none of the series shows discontinuities.

\subsection{Counterfactuals and official figures}
\label{sec:contrafactuales}

Table~\ref{tab:escenarios} collects the orbital lifetimes that result from replacing the measured
forcing with alternative scenarios while keeping the ballistic coefficient fixed. Under the
environmental scenario derived from the forecast available during design, the satellite would have
flown 3.83 years, a factor of $\times1.52$ more than the 2.53 simulated with the measured indices; a
\SI{10}{\percent} variation in the coefficient moves that factor between $1.46$ and $1.55$. The 2019
scenario alters not only \Fdiez{} but also the geomagnetic activity, which is derived
climatologically from the flux itself, so that the $\times1.52$ measures the effect of the complete
environment and not that of the flux difference in isolation. Under an average cycle of
\SI{120}{\sfu} it would have reached 5.36 years, compatible with the at least five declared for the
mission.

\begin{table}[htbp]
\centering
\caption{Simulated orbital lifetime of FACSAT-2 with
$B_{\mathrm{eff}}=\SI{72.3}{\kilo\gram\per\meter\squared}$ according to the assumed forcing, counted
from the first epoch of the archive. The scenarios derived from the 2019 forecast and from the
average cycle modify \Fdiez{} and the geomagnetic activity tied to it simultaneously; only the
second row isolates the geomagnetic component.}
\label{tab:escenarios}
\small
\begin{tabular}{@{}lrl@{}}
\toprule
\textbf{Scenario} & \textbf{Lifetime [yr]} & \textbf{Comment} \\
\midrule
Measured indices 2023--2025 & 2.53 & re-entry observed on 27 Oct 2025 \\
Quiet geomagnetism ($a_p = 6$) & 2.80 & $+102$ d; 2.58 with $a_p=8$ and 3.20 with $a_p=4$ \\
Scenario derived from the 2019 forecast & 3.83 & shortening factor $\times1.52$ \\
Average cycle (\SI{120}{\sfu}) & 5.36 & compatible with the at least five declared \\
Permanent solar minimum (\SI{70}{\sfu}) & 18.1 & \\
\bottomrule
\end{tabular}
\end{table}

\begin{figure}[htbp]
\centering
\includegraphics[width=0.98\textwidth]{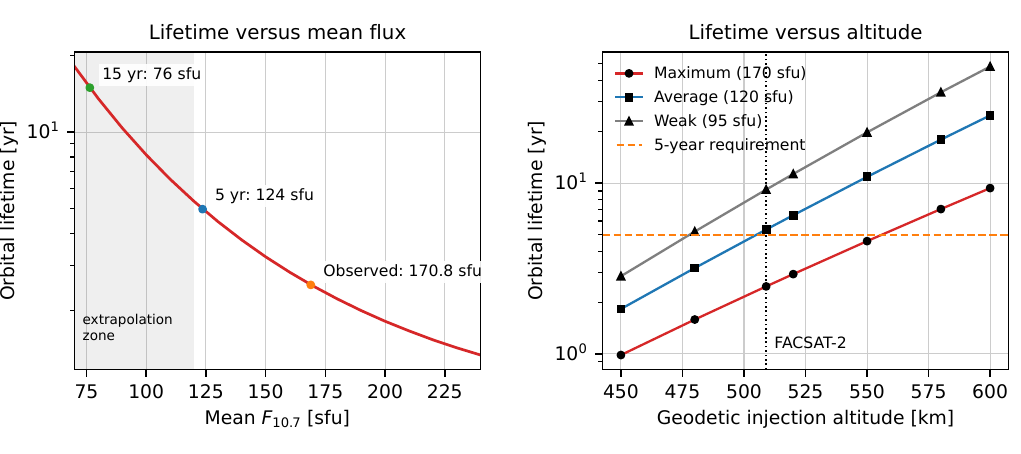}
\caption{Left, orbital lifetime from the injection orbit as a function of the mean flux, with the
values implicit in the two official figures and the one actually observed. Right, orbital lifetime
against injection altitude under three flux regimes.}
\label{fig:lifecurve}
\end{figure}

The two official figures admit the same treatment by inversion (Figure~\ref{fig:lifecurve}). The
nominal extrapolation places at some \SI{76}{\sfu} the mean flux needed to reach fifteen years from
the real orbit, a threshold that the comparison with FACSAT-1 suggests could shift towards
\SIrange{90}{100}{\sfu}; in any case these are values characteristic of a deep solar minimum.
Interpreting the five years of service life as a minimum requirement of orbital persistence, the
model assigns them \SI{124}{\sfu}, close to the mean of a typical cycle. Note that the model
determines which flux would be compatible with five years of persistence, not that the institutional
estimate was computed assuming that flux: the operational service life depends in addition on power,
radiation degradation and attitude control. The satellite flew in \SI{170.8}{\sfu}. The service-life
estimate was defensible; the orbital-lifetime one required a prolonged solar minimum even with the
most favourable threshold.

\begin{figure}[htbp]
\centering
\includegraphics[width=0.78\textwidth]{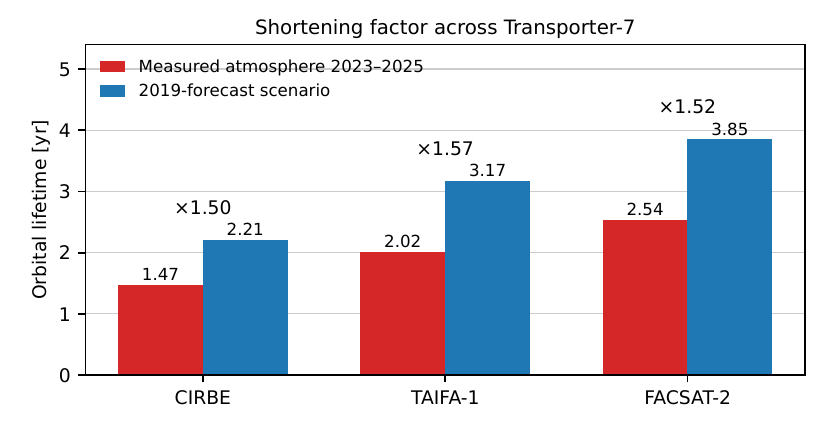}
\caption{Observed and counterfactual orbital lifetime of three Transporter-7 objects subjected to
the same measured atmosphere. The counterfactual corresponds to the environmental scenario derived
from the official 2019 forecast. The ballistic coefficients of CIRBE and TAIFA-1 are fitted to their
re-entry dates, so that the comparison is a consistency check and not an independent measurement.
The lifetimes in this figure are counted in the deployment--re-entry convention, which for FACSAT-2
gives 2.54 and 3.85 years, whereas Table~\ref{tab:escenarios} uses the lifetime simulated from the
first epoch of the archive and gives 2.53 and 3.83; both pairs describe the same quantity in
different conventions, as Section~S1 details, and yield the same factor $\times1.52$.}
\label{fig:companeros}
\end{figure}

The launch companions offer a consistency check across objects
(Figure~\ref{fig:companeros}). Subjected to the same measured atmosphere, CIRBE and TAIFA-1 would
have flown $\times1.50$ and $\times1.57$ longer under the 2019 scenario, factors very close to that
of FACSAT-2 despite their ballistic coefficients differing by almost a factor of two. It would be a
mistake to present this as independent proof: the three counterfactuals share an atmospheric model
and the coefficients of CIRBE and TAIFA-1 are fitted to their re-entry dates, and some similarity is
expected by construction. What the exercise establishes is that a single environment coherently
describes three vehicles of very different properties. The complementary experiment, which changes
only the launch date, indicates that FACSAT-2 launched in November 2018 would have flown 5.32 years
and FACSAT-1 launched in April 2023 only 1.60.

\subsection{Response to the geomagnetic storms}
\label{sec:tormentas}

\begin{figure}[htbp]
\centering
\includegraphics[width=0.98\textwidth]{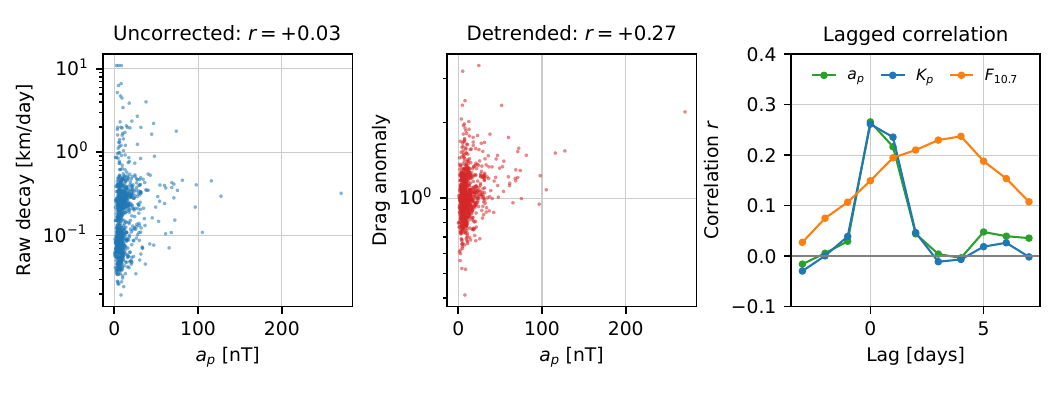}
\caption{Left, raw decay rate against $a_p$, whose correlation is null because the series is
dominated by the decrease of the altitude. Centre, the same relation for the drag anomaly. Right,
lagged correlation for $a_p$, $K_p$ and \Fdiez{}.}
\label{fig:correlaciones}
\end{figure}

The correlation between the raw decay rate and any index turns out to be misleading, and even
negative in sign for \Fdiez{}, because the dominant signal in that series is the monotonic decrease
of the altitude and not the forcing. Once the trend is removed through the anomaly, the correlations
become positive: $r = +0.27$ for $a_p$, $+0.26$ for $K_p$, $-0.26$ for the daily minimum of Dst and
$+0.15$ for \Fdiez{} (Figure~\ref{fig:correlaciones}). Correcting for the effective sample size,
which is reduced to between 632 and 397 independent observations depending on the index, the
probabilities fall between $1\times10^{-11}$ and $3\times10^{-3}$, and the block bootstrap yields
$p = 0.0002$ for the first three and $p = 0.012$ for \Fdiez{}: the geomagnetic correlations are
clearly significant and that of \Fdiez{} more modestly so, several orders of magnitude above what
the naive test would suggest.

The lag analysis reveals two distinct response times. Geomagnetic activity acts almost immediately,
with a maximum of $r = +0.27$ on the same day and values indistinguishable from zero from the second
day onwards, whereas \Fdiez{} reaches its maximum correlation some three days later, with
$r = +0.23$. The asymmetry is compatible with the thermal inertia of the thermosphere in the face of
auroral energy deposition, which heats it within hours, although the lag should not be taken as a
direct measurement of the thermal response time, because \Fdiez{} is an indirect proxy for the
ultraviolet irradiance. None of the negative lags shows any signal, and repeating the analysis with
smoothing windows of five to fifteen days leaves the maximum at the same lag, so it is not an
artefact of the filtering.

\begin{figure}[htbp]
\centering
\includegraphics[width=0.98\textwidth]{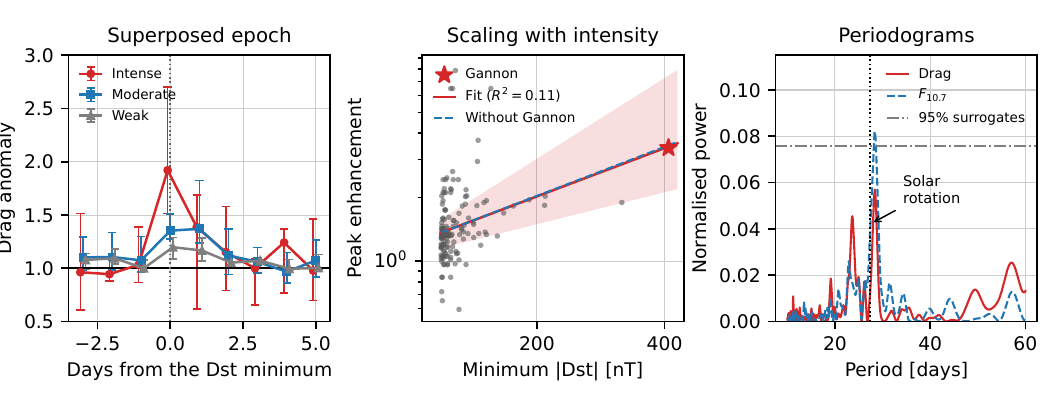}
\caption{Left, superposed epoch of the drag anomaly for three storm classes. Centre, peak
enhancement against storm intensity with the exponential fit. Right, periodograms of the drag
anomaly and of the observed \Fdiez{}, both peaking at $28.3\pm1.0$ days.}
\label{fig:tormentas}
\end{figure}

Table~\ref{tab:tormentas} and Figure~\ref{fig:tormentas} summarise the superposed epoch over the 149
storms. The six intense ones produce a median drag enhancement of $\times1.92$ on the day of the
minimum, with a \SI{95}{\percent} bootstrap interval between \num{1.51} and \num{2.70} that reflects
the small size of the sample, and a return to the background in two or three days; the moderate ones
give $\times1.35$ and the weak ones $\times1.20$; the ordering by intensity holds even though the
individual amplitudes are less well determined. An exponential fit gives
$1.20\,\exp(0.261\,|\mathrm{Dst}|/100)$, which survives the exclusion of the extreme event but
explains only \SI{11}{\percent} of the variance. Repeating the analysis after merging the minima
separated by less than 24, 48 and 72 hours, the six intense ones are unaffected and the moderate
ones barely change, whereas the weak ones are reduced from one hundred and fifteen to thirty-eight
and drop from $\times1.20$ to $\times1.07$: the two upper classes are robust against the clustering
criterion, the enhancement of the weak class depends on it.

\begin{table}[htbp]
\centering
\caption{Superposed epoch of the drag anomaly around the day of the Dst minimum, median by storm
class. In brackets, the \SI{95}{\percent} bootstrap interval of the median on the day of the
minimum.}
\label{tab:tormentas}
\small
\begin{tabular}{@{}lrrrrrr@{}}
\toprule
\textbf{Class} & \textbf{n} & \textbf{$-2$ d} & \textbf{$-1$ d} & \textbf{day 0} & \textbf{$+1$ d} & \textbf{$+2$ d} \\
\midrule
Intense (Dst $< \SI{-150}{\nano\tesla}$)  & 6   & 0.94 & 1.03 & 1.92 [1.51--2.70] & 1.39 & 1.15 \\
Moderate ($-150$ to \SI{-80}{\nano\tesla}) & 28  & 1.11 & 1.07 & 1.35 [1.28--1.51] & 1.37 & 1.12 \\
Weak ($-80 < \mathrm{Dst} \le \SI{-50}{\nano\tesla}$) & 115 & 1.09 & 1.00 & 1.20 [1.09--1.29] & 1.17 & 1.05 \\
\bottomrule
\end{tabular}
\end{table}

The Gannon superstorm, broken down at hourly resolution in Section~S7 and Figure~3S, is the extreme
case: the Dst minimum of \SI{-406}{\nano\tesla} of 11 May 2024 coincides on the same day with the
peak of $B^{*}$ and with the anomaly reaching $\times3.38$, with a return to the background in about
five days. That value falls to $\times1.29$ with a $\pm7$-day window, and should not be understood
as a robust physical amplitude. The third panel of Figure~\ref{fig:tormentas} adds evidence
compatible with a solar rotational modulation: the periodograms of the anomaly and of \Fdiez{} reach
their maximum at the same period, 28.3 days, compatible with the synodic rotation. Against random
permutations the false-alarm probability is below \num{0.002}, but against phase surrogates the peak
ceases to be distinguishable from the background ($p = 0.79$); what is informative is not the
isolated peak but the fact that both maxima coincide. One final caveat remains, namely that
attributing the re-entry to the last storm on the calendar would be a mistake, since the
perturbations of October 2025 \citep{swpc_alerts2025} found the satellite already below
\SI{300}{\kilo\meter}, so that they could shift the date by days but not by months.

Separating the geomagnetic contribution requires first fixing what is meant by a quiet atmosphere.
The nominal level adopted is $a_p = \SI{6}{\nano\tesla}$, equivalent to $K_p = 1.7$
\citep{matzka2021}, following the conventional threshold for a geomagnetically quiet day,
$K_p \le 2$, and it coincides with what the series itself offers as a typical quiet level: of the
924 days covered, the 498 whose mean $K_p$ did not exceed that threshold have a median $a_p$ of
\SI{5.6}{\nano\tesla}. The two extremes of the sensitivity test admit the same empirical reading;
restricting the quiet class to the 188 days on which $K_p$ did not exceed the threshold in any of
its eight three-hourly intervals, the median $a_p$ falls to \SI{3.8}{\nano\tesla}, so that
$a_p = \SI{4}{\nano\tesla}$ amounts to postulating the strictest definition of sustained calm for
almost three years, whereas \SI{8}{\nano\tesla} is very close to the median $a_p$ of all days,
\SI{8.1}{\nano\tesla}, and barely suppresses any activity.

With the nominal level, repeating the simulation with the real flux but with geomagnetic activity at
quiet conditions raises the orbital lifetime from 2.53 to 2.80 years, that is, it adds 102 days,
\SI{22}{\percent} of the total shortening of 1.30 years with respect to the design scenario. That
fraction is markedly sensitive to the threshold: with $a_p = \SI{4}{\nano\tesla}$ it returns 245
days, \SI{52}{\percent}, and with $a_p = \SI{8}{\nano\tesla}$ only 17, barely \SI{4}{\percent}. A
second route, independent of that definition, integrates the drag anomaly and attributes around
\SI{15}{\percent} of the altitude loss to enhancements faster than the 81-day window. The two
figures do not measure the same thing, and their difference is informative: geomagnetic activity
produces not only peaks of one or two days but also an elevated background during the active
periods, which the anomaly method assigns to the baseline and the counterfactual does count. Taken
together they place the geomagnetic contribution at around one fifth of the shortening under the
nominal scenario, with the caveat that the sensitivity to the threshold admits values between
\SI{4}{\percent} and \SI{52}{\percent} and that it is therefore the worst-determined quantity in
this work. Under the nominal scenario the sustained solar forcing clearly dominates; at the
$a_p = \SI{4}{\nano\tesla}$ extreme both contributions turn out to be comparable.

\subsection{Distribution of the orbital lifetime against solar scenarios}
\label{sec:percentiles}

The preceding scenarios are deterministic. A more useful question for design is a different one:
given that the phase and the amplitude of the cycle at the moment of launch are not known in
advance, what distribution of orbital lifetimes should be expected? We simulated \num{2000}
realisations per injection altitude, sampling the cycle amplitude uniformly between 115 and 285
smoothed sunspots, a range that spans the maxima of cycles 18 to 25, and the launch phase uniformly
within an eleven-year cycle; the trajectories that survive more than one cycle chain successive
cycles by drawing a new amplitude for each. The ensemble incorporates the uncertainty in the
amplitude and phase of the solar cycle but not the stochastic variability of the storms, since $a_p$
is tied deterministically to the flux, and Table~2S of the supplementary material represents a
distribution conditioned on the adopted ensemble and not a complete distribution of orbital risk.

From the real injection orbit the median orbital lifetime of the ensemble is 4.6 years, with 5th and
95th percentiles of 1.4 and 8.4, and only \SI{45}{\percent} of the realisations reach five years of
persistence; the 2.54 years observed fall in the 24th percentile, in the lower quartile but not in
the extreme tail. Raising the injection to \SI{560}{\kilo\meter} takes that fraction to
\SI{79}{\percent} and at \SI{580}{\kilo\meter} to \SI{89}{\percent}, which is the natural way of
translating a persistence requirement into a design altitude. These figures inherit the limitations
of the model, in particular the extrapolation towards low fluxes, and the uniform sampling of
amplitudes is a deliberately neutral choice.

%%%%%%%%%%%%%%%%%%%%%%%%%%%%%%%%%%%%%%%%%%%%%%%%%%%%%%
\section{Discussion}
\label{sec:discusion}

The possible explanations admit an ordering, and only some survive the confrontation with the data.
That of a solar forcing stronger than foreseen is supported and dominant, with a factor of
$\times1.52$ of which the greater part corresponds to sustained heating. That of an injection lower
than announced is resolved: the discrepancy of eleven kilometres is a matter of altitude convention
and the institutional figures are consistent with a nominal injection. That of a late loss of
attitude control finds no support, since it would appear as an abrupt and permanent jump of the
apparent ballistic coefficient, and the observed excursions, which reach a factor of two during the
first half of 2025, are reversible and coincide with the decline of the solar flux; the elements do
not, however, allow a gradual transition of operating mode that altered the projected area without a
detectable discontinuity to be excluded. As for fragmentation and manoeuvres there are no signs
either: the eccentricity decreases monotonically without jumps, the semi-major axis presents no
increases, the fragmentation-event counts record none attributable to this object
\citep{mcdowell2026} and the platform had no propulsion.

One hypothesis remains that the data bound but do not close. The effective area of
\SI{0.047}{\meter\squared} falls somewhat below the \SI{0.055}{\meter\squared} of a 6U in random
tumbling, which is consistent with a stabilised platform that would have preferentially presented a
smaller cross-section, without the data allowing it to be demonstrated that this configuration was
maintained throughout the mission. The dominant uncertainty is not the mass but the drag coefficient
and the absolute density level: varying $C_D$ between \num{2.0} and \num{2.4} moves the area between
\num{0.043} and \SI{0.052}{\meter\squared}, still below free tumbling, whereas a bias of
$\pm\SI{30}{\percent}$ in the density takes it to \SIrange{0.036}{0.067}{\meter\squared} and at its
upper end would reverse the reading. The interpretation is reinforced by the fact that the vehicle
had attitude control with reaction wheels, magnetorquers and star tracking
\citep{fac_facsat2,rincon2023} and by the fact that a camera of \SI{4.75}{\meter} resolution only
performs if the pointing is stable.

The comparison between the two Colombian missions misleads if it is made naively. FACSAT-1 lasted
4.51 years and FACSAT-2, larger and technologically superior, 2.54; but the former flew during the
deep minimum between cycles 24 and 25, with a mean flux close to \SI{97}{\sfu} against the
\SI{170.8}{\sfu} of the latter, a difference which, according to the sensitivity of the corrected
model, a factor of $3.15$ per \SI{100}{\sfu}, corresponds to a thermospheric density somewhat more
than twice as large. Applying the model with the coefficient of each one, FACSAT-1 launched in April
2023 would have lasted 1.60 years and FACSAT-2 launched in November 2018, 5.32. The management
implication is direct: any indicator comparing missions separated by several years should be
normalised by the solar forcing of the period, or it will end up rewarding or punishing teams for
the phase of the cycle that fell to them.

FACSAT-1 additionally provides the most eloquent precedent of this work. In 2021, with a little more
than two years of elements and a high-fidelity propagator, \citet{portilla2021} placed its re-entry
in the first half of 2030; the object re-entered on 3 June 2023, almost seven years earlier, which
amounts to overestimating the orbital lifetime by a factor of $2.5$. The method was the correct one:
what failed was the one variable that nobody controls, because an atmosphere fed with the activity
of the 2016--2021 period, an exceptionally deep minimum, could not anticipate the cycle that came
afterwards. That the same mechanism should have invalidated equally a careful academic prediction
and an institutional estimate suggests that the problem does not lie in the rigour of whoever
computes but in treating future solar activity as a datum and not as a distribution.

That same article offers a partial external check of the behaviour of the model under low-flux
conditions. \citet{portilla2021} measured decay rates of \SI{8}{\meter\per\day} and
\SI{18}{\meter\per\day} at two epochs of the solar minimum, whereas our model returns
\SI{13}{\meter\per\day} and \SI{20}{\meter\per\day}: it overestimates the drag by around
\SI{60}{\percent} during the minimum and by \SI{13}{\percent} during the ascending phase. This is
not surprising, because the correction of Equation~\eqref{eq:correccion} was fitted with fluxes
between \num{135} and \SI{253}{\sfu}, but it has a practical consequence: the low-flux scenarios of
Table~\ref{tab:altura} and the value of \SI{76}{\sfu} associated with the fifteen years should be
read as conservative bounds, and the flux that would make that official figure plausible would rise
towards \SIrange{90}{100}{\sfu}, still far below the \SI{170.8}{\sfu} experienced.

The decay also had a little-discussed side effect on the payload. The ground sampling distance
scales linearly with altitude: the \SI{4.75}{\meter} nominal at \SI{500}{\kilo\meter}
\citep{simera2021} became some \SI{3.3}{\meter} at \SI{350}{\kilo\meter}: the mission gained
resolution while it lost altitude, but it lost swath width in the same proportion, from
\SI{19.4}{\kilo\meter} to some \SI{13.6}{\kilo\meter}, degrading the coverage of the territory
within the four-day window that the requirements set \citep{rincon2025}. A third effect usually goes
unnoticed: the nodal precession rate departed from the ideal Sun-synchronous value, from
\ang{0.9887} per day at the beginning to \ang{1.1544} at the end, accumulating \ang{26.7} equivalent
to 1.8 hours of drift in the local solar time of the node. The scientific capability can therefore
degrade well before the vehicle re-enters.

As for design, Table~\ref{tab:altura} quantifies the altitude trade-off: a five-year requirement
demands injection at around \SI{560}{\kilo\meter} if the flight takes place at a solar maximum,
whereas \SI{510}{\kilo\meter} would suffice under an average cycle. The choice is not free on a
rideshare mission and it also clashes with the debris-mitigation criterion: the limit of twenty-five
years of residual persistence after end of mission set out in the IADC guidelines \citep{iadc2021}
and in the ISO 24113 standard \citep{iso24113}, with a trend towards shorter deadlines such as the
five years adopted by the FCC in 2022 \citep{fcc2022}. The comparison requires an intermediate step,
because the table collects the total lifetime from injection and the standard regulates residual
persistence, which for a passive vehicle is the total lifetime minus the operational phase. Adopting
five years of operational phase, a vehicle with the coefficient of FACSAT-2 injected at
\SI{600}{\kilo\meter} would leave 19.9 years of residual persistence under an average cycle and 43.2
under a weak one; at \SI{580}{\kilo\meter}, 13.0 and 29.1. Non-compliance depends not only on the
altitude but also on the phase of the cycle that the mission goes through.

The alternative is propulsion. With the measured forcing, maintaining the injection orbit during the
\num{2.53} years of the archive would have cost \SI{36}{\meter\per\second}. Extending the
calculation to five years requires the assumption to be stated: the 2019 scenario requires
\SI{41}{\meter\per\second} and an average cycle \SI{29}{\meter\per\second}, because both include the
descending branch, whereas maintaining the pace of the maximum for five years would lead to some
\SI{70}{\meter\per\second}. That budget must be translated into the quantity that characterises a
thruster, because velocity increment and total impulse are not dimensionally the same: for
\SI{7.49}{\kilo\gram} and to first approximation, \SI{30}{\meter\per\second} is equivalent to some
\SI{225}{\newton\second} and \SI{70}{\meter\per\second} to some \SI{524}{\newton\second}, figures
within reach of miniaturised electric propulsion for 6U platforms \citep{krejci2018}. There is
moreover a third lever that costs no mass: the phase of the solar cycle at the moment of launch is a
mission parameter, not a meteorological datum.

\begin{table}[htbp]
\centering
\caption{Total orbital lifetime from injection of a 6U CubeSat with
$B_{\mathrm{eff}}=\SI{72.3}{\kilo\gram\per\meter\squared}$ according to injection altitude, in the
orbital geodetic convention, and the mean solar flux regime. The three regimes are illustrative
scenarios of constant mean flux and not averages of a particular historical sample:
\SI{170}{\sfu} approximately reproduces the mean value of the mission, \SI{120}{\sfu} the order of
magnitude of a typical complete cycle and \SI{95}{\sfu} a distinctly weak cycle. To obtain the
residual persistence after end of mission, which is the regulated quantity, the duration of the
operational phase must be subtracted.}
\label{tab:altura}
\small
\begin{tabular}{@{}lrrr@{}}
\toprule
\textbf{Injection altitude} & \textbf{Solar max.\ (\SI{170}{\sfu})} & \textbf{Average cycle (\SI{120}{\sfu})} & \textbf{Weak cycle (\SI{95}{\sfu})} \\
\midrule
\SI{450}{\kilo\meter} & 1.0 yr & 1.8 yr & 2.9 yr \\
\SI{480}{\kilo\meter} & 1.6 yr & 3.2 yr & 5.2 yr \\
\textbf{\SI{509}{\kilo\meter}} (FACSAT-2) & \textbf{2.5 yr} & \textbf{5.4 yr} & \textbf{9.2 yr} \\
\SI{520}{\kilo\meter} & 2.9 yr & 6.5 yr & 11.3 yr \\
\SI{550}{\kilo\meter} & 4.6 yr & 10.9 yr & 19.8 yr \\
\SI{580}{\kilo\meter} & 7.1 yr & 18.0 yr & 34.1 yr \\
\SI{600}{\kilo\meter} & 9.4 yr & 24.9 yr & 48.2 yr \\
\bottomrule
\end{tabular}
\end{table}

%%%%%%%%%%%%%%%%%%%%%%%%%%%%%%%%%%%%%%%%%%%%%%%%%%%%%%
\section{Limitations}
\label{sec:limitaciones}

The main limitation is the degeneracy between the ballistic coefficient and the absolute density
level: what orbital tracking constrains is the ratio
$\rho/B_{\mathrm{eff}} = \rho\,C_D A/m$, and a bias in the density translates directly into a bias
of the same sign in the inferred coefficient. Sampling with NRLMSISE-00 breaks that degeneracy only
to the extent that the accuracy of that model is accepted, its typical uncertainty being of the
order of \SI{20}{\percent} \citep{emmert2015}, so that the effective area inherits that uncertainty
and leaves the discussion about the attitude of the vehicle open.

The simplified model is corrected within the range of conditions observed, so the scenarios with
fluxes between \SI{70}{\sfu} and \SI{120}{\sfu} constitute an extrapolation that is difficult to
bound without a control mission that has flown under those conditions. Nor does it explicitly
include the seasonal variation or the dependence on local solar time; the latter was approximately
constant during the first part of the mission, as corresponds to a Sun-synchronous orbit, but the
acceleration of the nodal precession produced a drift of 1.8 hours concentrated at the end of the
flight. No archive of elements was available for FACSAT-1, so its treatment rests on its published
insertion orbit \citep{portilla2021}, on the re-entry date and on a parametric reconstruction of the
forcing; the same applies to the masses and areas of the comparison objects. Finally, the elements
used are mean in the sense of SGP4 theory \citep{vallado2006} and inherit its conventions,
particularly during the last weeks of flight, when $B^{*}$ ceases to have physical meaning.

%%%%%%%%%%%%%%%%%%%%%%%%%%%%%%%%%%%%%%%%%%%%%%%%%%%%%%
\section{Conclusions}
\label{sec:conclusiones}

The joint analysis of the archive of orbital elements and of the measured indices shows that the
premature re-entry of FACSAT-2 requires no extraordinary explanations, but follows from a solar
cycle considerably more intense than forecast acting on a vehicle that behaved as its geometry would
lead one to expect.

\begin{enumerate}
  \item The inversion of the observed drag yields
        $B_{\mathrm{eff}} = \SI{72.3}{\kilo\gram\per\meter\squared}$, with a statistical uncertainty
        of \SI{6}{\percent} and a systematic one of the order of \SI{20}{\percent} dominated by the
        density model, a value that for the nominal mass implies an effective area of
        \SI{0.047}{\meter\squared}, compatible with the geometry of a 6U. With that coefficient
        fixed, the integration reproduces the \num{922.7} days of the archive with a root-mean-square
        error of \SI{7.4}{\kilo\meter}, and calibrated only with 2023 and 2024 it places the
        re-entry with a 28-day out-of-sample error.
  \item The \Fdiez{} flux averaged \SI{170.8}{\sfu} against the \SI{\Fesc}{\sfu} that the
        environmental scenario derived from the official 2019 forecast returns over the same window.
        That scenario, which modifies both the flux and the geomagnetic activity climatologically
        tied to it, produces an orbital lifetime $\times1.52$ longer. The two launch companions yield
        $\times1.50$ and $\times1.57$, which constitutes a consistency check across objects and not
        an independent measurement.
  \item Replacing the observed geomagnetic activity with quiet conditions
        ($a_p = \SI{6}{\nano\tesla}$) adds 102 days of lifetime, \SI{22}{\percent} of the
        shortening, whereas integrating the drag anomaly attributes around \SI{15}{\percent} of the
        altitude loss to the fast enhancements. Both routes place the geomagnetic contribution at
        around one fifth of the shortening, although it is the worst-determined quantity in this
        work: the sensitivity to the level adopted as quiet admits values between \SI{4}{\percent}
        and \SI{52}{\percent}. The remainder corresponds to the sustained solar forcing, which
        clearly dominates under the nominal scenario. The effect of the storms is intense but brief,
        with a median enhancement of $\times1.92$ on the day of the Dst minimum for the intense
        events and a return to the background in two or three days.
  \item The model allows the two official figures to be inverted and the mean flux with which they
        would be compatible to be determined, without this implying that the institutional estimate
        was computed under that assumption: some \SI{124}{\sfu} for five years of orbital
        persistence, which is defensible, and some \SI{76}{\sfu} for the fifteen years, a threshold
        that the comparison with FACSAT-1 suggests could shift towards \SIrange{90}{100}{\sfu} and
        that in any case required a prolonged solar minimum.
  \item The scientific capability of an Earth-observation mission can degrade before re-entry: the
        camera swath narrowed from \SI{19.4}{\kilo\meter} to some \SI{13.6}{\kilo\meter} and the
        nodal precession accumulated 1.8 hours of drift in the local solar time. Defining an
        end-of-service date would require a quantitative operational criterion that should be set at
        the design stage.
  \item For similar configurations, an injection of the order of \SI{560}{\kilo\meter}, with margin
        towards \SI{580}{\kilo\meter}, or a propulsive capability of approximately
        \SIrange{30}{70}{\meter\per\second} depending on the solar scenario, reaching some
        \SI{70}{\meter\per\second} in the conservative case of a sustained maximum, would sustain
        five years of orbit. In an ensemble of \num{2000} solar-cycle scenarios, from the real orbit
        the median lifetime is 4.6 years with 5th and 95th percentiles of 1.4 and 8.4, and only
        \SI{45}{\percent} of the realisations reach five years, against \SI{79}{\percent} at
        \SI{560}{\kilo\meter} and \SI{89}{\percent} at \SI{580}{\kilo\meter}.
\end{enumerate}

%%%%%%%%%%%%%%%%%%%%%%%%%%%%%%%%%%%%%%%%%%%%%%%%%%%%%%
\section*{Acknowledgements}
\addcontentsline{toc}{section}{Acknowledgements}

The authors thank the maintainers of the public space-tracking catalogues, whose availability made
this analysis possible. The structure of the manuscript made use of the assistance of an artificial
intelligence language model; the authors are responsible for the verification of all the results
presented. This work has not been funded or endorsed by the Military Forces of Colombia.

%\section*{Data and code availability}
%\addcontentsline{toc}{section}{Data and code availability}

%The input data are public: the archive of orbital elements of object NORAD 56205 comes from the
%Space-Track catalogue \citep{spacetrack2025}, with CelesTrak as a secondary source, and the solar
%and geomagnetic indices from the OMNI2 database of the Space Physics Data Facility
%\citep{king2005}. The derived material supporting the results will be deposited in a public
%repository with a permanent identifier, and includes the processed archive of elements, the daily
%series of altitude, decay rate and orbital mean density, the catalogue of 149 storms, the
%NRLMSISE-00 evaluation code through \texttt{pymsis} \citep{pymsis2022}, the orbital simulator and
%the scripts for the figures and the statistical procedures. Section~S4 of the supplementary material
%states the number of realisations and the randomisation procedures; the exact seeds are provided in
%the code.

%%%%%%%%%%%%%%%%%%%%%%%%%%%%%%%%%%%%%%%%%%%%%%%%%%%%%%
\clearpage
\renewcommand{\bibname}{References}
%% references.tex -- bibliography in APA format (7th edition)
%% Written by hand because the LaTeX installation includes neither apacite nor
%% biblatex-apa. The keys are natbib-compatible: \citep and \citet behave exactly
%% as they would with a .bbl generated by bibtex. The file refs.bib is kept in the
%% package as the data source, but the bibliography that is typeset is this one.
%% Titles originally published in Spanish are given in the original language with
%% an English translation in square brackets, as APA prescribes.


\begin{thebibliography}{}
\setlength{\itemsep}{4pt}

\bibitem[Bates(1959)]{bates1959}
Bates, D. R. (1959). Some problems concerning the terrestrial atmosphere above about the 100 km
level. \emph{Proceedings of the Royal Society A}, \emph{253}(1275), 451--462.
\url{https://doi.org/10.1098/rspa.1959.0207}

\bibitem[C\'ardenas et~al.(2023)]{cardenas2023}
C\'ardenas, L., Guti\'errez, E., Chac\'on, W., Plata, I., \& Moreno, C. (2023). Investigaci\'on para
la implementaci\'on de una segunda carga \'util al sistema satelital FACSAT-2, para el an\'alisis de
gases de efecto invernadero [Research towards the implementation of a second payload on the FACSAT-2
satellite system for the analysis of greenhouse gases]. In \emph{Gesti\'on tecnol\'ogica e
innovaci\'on en el campo aeroespacial} (Chap.~5, pp.~127--170). Sello Editorial Escuela Militar de
Aviaci\'on ``Marco Fidel Su\'arez''.

\bibitem[Clette(2021)]{clette2021}
Clette, F. (2021). Is the F10.7cm--sunspot number relation linear and stable?
\emph{Journal of Space Weather and Space Climate}, \emph{11}, 2.
\url{https://doi.org/10.1051/swsc/2020071}

\bibitem[Corredor and Benavides(2019)]{corredor2019}
Corredor, G., \& Benavides, E. (2019). Transferencia de tecnolog\'ia y desarrollo de capacidades para
el programa espacial colombiano mediante peque\~nos sat\'elites [Technology transfer and capacity
building for the Colombian space programme through small satellites]. \emph{Revista de Tecnolog\'ia
Aeron\'autica}, \emph{29}, 8--19.

\bibitem[Departamento Nacional de Planeaci\'on(2020)]{conpes3983}
Departamento Nacional de Planeaci\'on. (2020). \emph{Documento Conpes 3983: Pol\'itica de desarrollo
espacial} [Conpes document 3983: Space development policy]. Government of Colombia.
\url{https://colaboracion.dnp.gov.co/CDT/Conpes/Econ\%C3\%B3micos/3983.pdf}

\bibitem[Emmert(2015)]{emmert2015}
Emmert, J. T. (2015). Thermospheric mass density: A review. \emph{Advances in Space Research},
\emph{56}(5), 773--824. \url{https://doi.org/10.1016/j.asr.2015.05.038}

\bibitem[Fang et~al.(2022)]{fang2022}
Fang, T.-W., Kubaryk, A., Goldstein, D., Li, Z., Fuller-Rowell, T., Millward, G., Singer, H. J.,
Steenburgh, R., Westerman, S., \& Babcock, E. (2022). Space weather environment during the SpaceX
Starlink satellite loss in February 2022. \emph{Space Weather}, \emph{20}(11), e2022SW003193.
\url{https://doi.org/10.1029/2022SW003193}

\bibitem[Ford(2026)]{inthesky}
Ford, D. (2026). \emph{Spacecraft information: FACSAT-2, CIRBE, TAIFA-1}. In-The-Sky.org.
\url{https://in-the-sky.org/spacecraft.php?id=56205}

\bibitem[Colombian Aerospace Force(2023)]{fac_facsat2}
Colombian Aerospace Force. (2023). \emph{FACSAT-2 (SAT-Chiribiquete)}. Poder Aeroespacial FAC.
\url{https://poderespacial.fac.mil.co/facsat-2}

\bibitem[Colombian Aerospace Force(2020)]{fac_factibilidad}
Colombian Aerospace Force. (2020). \emph{Estudio de factibilidad para el desarrollo del proyecto
FACSAT-2} [Feasibility study for the development of the FACSAT-2 project]. Centro de
Investigaci\'on en Tecnolog\'ias Aeroespaciales (CITAE).

%% ENTRY NOT CITED IN THE TEXT: it was used only in the commented-out paragraph of
%% the introduction. Uncomment if that paragraph is restored; otherwise delete.
%% \bibitem[Government of Colombia(2019)]{estrategia2050}
%% Government of Colombia. (2019). \emph{Estrategia 2050: Cambio clim\'atico y gesti\'on de riesgos}
%% [Strategy 2050: Climate change and risk management]. Ministry of Environment and Sustainable
%% Development.
%% \url{https://www.minambiente.gov.co/cambio-climatico-y-gestion-del-riesgo/estrategia-2050/}

\bibitem[FCC(2022)]{fcc2022}
Federal Communications Commission. (2022). \emph{Space innovation; Mitigation of orbital debris in
the new space age: Second report and order} (FCC 22-74, IB Docket Nos.~18-313 and 22-271).
\url{https://docs.fcc.gov/public/attachments/FCC-22-74A1.pdf}

\bibitem[Hathaway(2015)]{hathaway2015}
Hathaway, D. H. (2015). The solar cycle. \emph{Living Reviews in Solar Physics}, \emph{12}, 4.
\url{https://doi.org/10.1007/lrsp-2015-4}

\bibitem[IADC(2021)]{iadc2021}
Inter-Agency Space Debris Coordination Committee. (2021). \emph{IADC space debris mitigation
guidelines} (IADC-02-01, Rev.~3). \url{https://iadc-home.org/documents_public/view/id/172}

\bibitem[ISO(2023)]{iso24113}
International Organization for Standardization. (2023). \emph{Space systems --- Space debris
mitigation requirements} (ISO 24113:2023, 4th ed.).
\url{https://www.iso.org/standard/83494.html}

\bibitem[Jacchia(1971)]{jacchia1971}
Jacchia, L. G. (1971). \emph{Revised static models of the thermosphere and exosphere with empirical
temperature profiles} (SAO Special Report No.~332). Smithsonian Astrophysical Observatory.

\bibitem[Jagpal et~al.(2010)]{jagpal2010}
Jagpal, R., Quine, B., Chesser, H., Abrarov, S., \& Lee, R. (2010). Calibration and in-orbit
performance of the Argus 1000 spectrometer. \emph{Journal of Applied Remote Sensing}, \emph{4}(1),
049501. \url{https://doi.org/10.1117/1.3302405}

\bibitem[Jallad et~al.(2019)]{jallad2019}
Jallad, A.-H., Marpu, P., Abdul Aziz, Z., Al Marar, A., \& Awad, M. (2019). MeznSat: A 3U CubeSat for
monitoring greenhouse gases using short wave infra-red spectrometry: Mission concept and analysis.
\emph{Aerospace}, \emph{6}(11), 118. \url{https://doi.org/10.3390/aerospace6110118}

\bibitem[King and Papitashvili(2005)]{king2005}
King, J. H., \& Papitashvili, N. E. (2005). Solar wind spatial scales in and comparisons of hourly
Wind and ACE plasma and magnetic field data. \emph{Journal of Geophysical Research: Space Physics},
\emph{110}, A02104. \url{https://doi.org/10.1029/2004JA010649}

\bibitem[Krejci and Lozano(2018)]{krejci2018}
Krejci, D., \& Lozano, P. (2018). Space propulsion technology for small spacecraft.
\emph{Proceedings of the IEEE}, \emph{106}(3), 362--378.
\url{https://doi.org/10.1109/JPROC.2017.2778747}

\bibitem[King-Hele(1987)]{kinghele1987}
King-Hele, D. (1987). \emph{Satellite orbits in an atmosphere: Theory and applications}. Blackie.

\bibitem[Lomb(1976)]{lomb1976}
Lomb, N. R. (1976). Least-squares frequency analysis of unequally spaced data.
\emph{Astrophysics and Space Science}, \emph{39}(2), 447--462.
\url{https://doi.org/10.1007/BF00648343}

\bibitem[Lucas(2022)]{pymsis2022}
Lucas, G. (2022). \emph{pymsis: A Python wrapper for the NRLMSIS empirical atmospheric models}
[Software]. \url{https://doi.org/10.5281/zenodo.5348502}

\bibitem[Matzka et~al.(2021)]{matzka2021}
Matzka, J., Stolle, C., Yamazaki, Y., Bronkalla, O., \& Morschhauser, A. (2021). The geomagnetic Kp
index and derived indices of geomagnetic activity. \emph{Space Weather}, \emph{19}(5), e2020SW002641.
\url{https://doi.org/10.1029/2020SW002641}

\bibitem[McDowell(2026)]{mcdowell2026}
McDowell, J. C. (2026). \emph{Space activities in 2025} (Rev. 1.4). Jonathan's Space Report.
\url{https://planet4589.org/space/papers/space25.pdf}

\bibitem[N2YO(2025)]{n2yo56205}
N2YO. (2025). \emph{Technical details for satellite FACSAT-2 (NORAD 56205)}.
\url{https://www.n2yo.com/satellite/?s=56205}

\bibitem[NASA(2025)]{nasa_cirbe2025}
NASA. (2025, February 6). \emph{NASA CubeSat finds new radiation belts after May 2024 solar storm}.
\url{https://science.nasa.gov/science-research/heliophysics/nasa-cubesat-finds-new-radiation-belts-after-may-2024-solar-storm/}

\bibitem[NASA(2026)]{nasa_vanallen2026}
NASA. (2026). \emph{Re-entry of the Van Allen Probe A spacecraft}.
\url{https://science.nasa.gov/mission/van-allen-probes/}

\bibitem[NOAA SWPC(2019)]{swpc_cycle2019}
National Oceanic and Atmospheric Administration, Space Weather Prediction Center. (2019).
\emph{Solar Cycle 25 forecast update: NOAA/NASA/ISES panel}.
\url{https://www.swpc.noaa.gov/news/solar-cycle-25-forecast-update}

\bibitem[NOAA SWPC(2025)]{swpc_alerts2025}
National Oceanic and Atmospheric Administration, Space Weather Prediction Center. (2025).
\emph{Geomagnetic storm alerts and warnings, October and November 2025}.
\url{https://www.swpc.noaa.gov/products/notifications-timeline}

\bibitem[NOAA SWPC(2026)]{swpc_progression}
National Oceanic and Atmospheric Administration, Space Weather Prediction Center. (2026).
\emph{Solar cycle progression}. \url{https://www.swpc.noaa.gov/products/solar-cycle-progression}

\bibitem[Oliveira et~al.(2025)]{oliveira2025}
Oliveira, D. M., Zesta, E., \& Nandy, D. (2025). The 10 October 2024 geomagnetic storm may have caused
the premature reentry of a Starlink satellite. \emph{Frontiers in Astronomy and Space Sciences},
\emph{11}, 1522139. \url{https://doi.org/10.3389/fspas.2024.1522139}

\bibitem[Parker and Linares(2024)]{parker2024}
Parker, W. E., \& Linares, R. (2024). Satellite drag analysis during the May 2024 Gannon geomagnetic
storm. \emph{Journal of Spacecraft and Rockets}. \url{https://doi.org/10.2514/1.A36164}

\bibitem[Picone et~al.(2002)]{picone2002}
Picone, J. M., Hedin, A. E., Drob, D. P., \& Aikin, A. C. (2002). NRLMSISE-00 empirical model of the
atmosphere: Statistical comparisons and scientific issues. \emph{Journal of Geophysical Research:
Space Physics}, \emph{107}(A12), 1468. \url{https://doi.org/10.1029/2002JA009430}

\bibitem[Pi\~neros et~al.(2021)]{pineros2021}
Pi\~neros, J. O. M., Dos Santos, W. A., \& Prado, A. F. B. A. (2021). Analysis of the orbit lifetime of
CubeSats in low Earth orbits including periodic variation in drag due to attitude motion.
\emph{Advances in Space Research}, \emph{67}(2), 902--918.
\url{https://doi.org/10.1016/j.asr.2020.10.024}

\bibitem[Portilla(2012)]{portilla2012}
Portilla, J. G. (2012). La \'orbita del sat\'elite Libertad 1 [The orbit of the Libertad 1
satellite]. \emph{Revista de la Academia Colombiana de Ciencias Exactas, F\'isicas y Naturales},
\emph{36}(141), 491--500.

\bibitem[Portilla and Murcia Pi\~neros(2021)]{portilla2021}
Portilla Barbosa, J. G., \& Murcia Pi\~neros, J. O. (2021). Evoluci\'on orbital del sat\'elite
FACSAT-1 y estimaci\'on de su tiempo de reentrada [Orbital evolution of the FACSAT-1 satellite and
estimation of its re-entry time]. \emph{Ciencia y Poder A\'ereo}, \emph{16}(1), 6--17.
\url{https://doi.org/10.18667/cienciaypoderaereo.694}

\bibitem[Rinc\'on et~al.(2023)]{rincon2023}
Rinc\'on, S., C\'ardenas, J., Piraz\'an, K., Acero, I., Hurtado, R., \& Cort\'es, D. (2023). Critical
design of the FACSAT-2 mission CubeSat for the observation and analysis of the Colombian territory.
\emph{Revista UIS Ingenier\'ias}, \emph{22}(3), 69--86.
\url{https://doi.org/10.18273/revuin.v22n3-2023006}

\bibitem[Rinc\'on Urbina et~al.(2025)]{rincon2025}
Rinc\'on Urbina, S. R., C\'ardenas Espinosa, L. P., Castillo Sep\'ulveda, M., Pirazan Villanueva,
K. N., Guti\'errez Bossa, E. E., \& Z\'arate Luna, P. A. (2025). Implementaci\'on de un modelo de
vigilancia tecnol\'ogica para la selecci\'on de las cargas \'utiles de la misi\'on FACSAT-2
[Implementation of a technology-watch model for the selection of the payloads of the FACSAT-2
mission]. \emph{Ciencia e Ingenier\'ia Neogranadina}, \emph{35}(2), 13--26.
\url{https://doi.org/10.18359/rcin.7581}

\bibitem[Simera Sense(2021)]{simera2021}
Simera Sense. (2021). \emph{MultiScape100 CIS datasheet}.
\url{https://www.simera-sense.com/download/multiscape100-cis/}

\bibitem[SILSO(2026)]{silso}
Sunspot Index and Long-term Solar Observations. (2026). \emph{Sunspot number and long-term solar
observations}. Royal Observatory of Belgium. \url{https://www.sidc.be/SILSO/}

\bibitem[Thoth Technology(2018)]{thoth2018}
Thoth Technology Inc. (2018). \emph{Argus 2000 IR spectrometer owner's manual OG274001}
(Version 1.03).

\bibitem[U.S. Space Force(2025)]{spacetrack2025}
United States Space Force, 18th Space Defense Squadron. (2025). \emph{Two-line element sets for object
NORAD 56205 (FACSAT-2)} [Data set]. Space-Track.org. \url{https://www.space-track.org}

\bibitem[Vallado et~al.(2006)]{vallado2006}
Vallado, D. A., Crawford, P., Hujsak, R., \& Kelso, T. S. (2006). \emph{Revisiting Spacetrack Report
No.~3} (AIAA 2006-6753). AIAA/AAS Astrodynamics Specialist Conference.
\url{https://doi.org/10.2514/6.2006-6753}

\bibitem[Vargas-Dom\'inguez et~al.(2026)]{vargas2026}
Vargas-Dom\'inguez, S., Agudelo-Rueda, J. A., Buitrago-Casas, J. C., Guerrero-Caguasango, H. D., \&
Calvo-Mozo, B. (2026). Actividad solar y clima espacial: desaf\'ios cient\'ificos y oportunidades
para el desarrollo en Colombia [Solar activity and space weather: Scientific challenges and
opportunities for development in Colombia]. \emph{Revista de la Academia Colombiana de Ciencias
Exactas, F\'isicas y Naturales}, \emph{50}(194), 191--212.
\url{https://doi.org/10.18257/raccefyn.3286}

%% ENTRY NOT CITED IN THE TEXT: it was used only in the commented-out paragraph of
%% the introduction. Uncomment if that paragraph is restored; otherwise delete.
%% \bibitem[Villela et~al.(2019)]{villela2019}
%% Villela, T., Costa, C. A., Brand\~ao, A. M., Bueno, F. T., \& Leonardi, R. (2019). Towards the
%% thousandth CubeSat: A statistical overview. \emph{International Journal of Aerospace Engineering},
%% \emph{2019}, 5063145. \url{https://doi.org/10.1155/2019/5063145}

%% ENTRY NOT CITED IN THE TEXT: it was used only in the commented-out paragraph of
%% the introduction. Uncomment if that paragraph is restored; otherwise delete.
%% \bibitem[Zhao et~al.(2022)]{zhao2022}
%% Zhao, Q., Yu, L., Du, Z., Peng, D., Hao, P., Zhang, Y., \& Gong, P. (2022). An overview of the
%% applications of Earth observation satellite data: Impacts and future trends. \emph{Remote Sensing},
%% \emph{14}(8), 1863. \url{https://doi.org/10.3390/rs14081863}
%%
\end{thebibliography}
\end{document}

\thispagestyle{plain}

\begin{center}
{\LARGE\bfseries Supplementary material\par}
\vspace{0.8em}
{\large Premature re-entry of the FACSAT-2 nanosatellite:\\[2pt]
orbital decay and attribution to the maximum of solar cycle 25\par}
\vspace{1.0em}
{\large Santiago Vargas Dom\'inguez, \quad Mara Valentina Angel, \quad Brayan Nicolas Buitrago\par}
\vspace{0.7em}
{\small Observatorio Astron\'omico Nacional, Universidad Nacional de Colombia, Bogot\'a, Colombia\par}
\end{center}

\vspace{1.2em}

\noindent
This document gathers the details of convention, implementation and validation that underpin the
results of the main article, together with three complementary figures. References to equations,
sections, tables and figures without a prefix point to the main article.

\vspace{1.0em}

%%%%%%%%%%%%%%%%%%%%%%%%%%%%%%%%%%%%%%%%%%%%%%%%%%%%%%
\section{Altitude and time conventions}
\label{sup:convenciones}

The problem admits two definitions of altitude that differ by more than ten kilometres and whose
confusion produces apparent discrepancies. The \emph{mean spherical altitude} is defined as the
semi-major axis minus the Earth's equatorial radius, $a - R_E$, and equals \SI{497.0}{\kilo\meter}
at the first epoch of the archive. The \emph{mean orbital geodetic altitude} is obtained by
evaluating the altitude above the WGS84 ellipsoid at the positions propagated with SGP4 and
averaging along one revolution; it equals \SI{509.1}{\kilo\meter} at that same epoch. The difference
comes from the Earth's oblateness and from the fact that a nearly polar orbit spends most of its
time over high latitudes, where the radius of the ellipsoid is smaller.

The second definition is the one that determines the density actually encountered. The
\SI{500}{\kilo\meter} of the payload specification \citep{simera2021,rincon2025} and the
\SI{508}{\kilo\meter} of the institutional documentation are consistent with this second definition
and, in any case, with a nominal injection close to \SI{500}{\kilo\meter}: the apparent discrepancy
of eleven kilometres with the altitude derived from the elements indicates no anomaly. It cannot be
asserted with certainty which convention the mission documentation uses, since it could be applying
a simplified orbital definition.

The article uses both conventions according to the quantity concerned, and each figure and table
states which one it uses. The series derived directly from the elements, such as the altitudes of
Figure~1 and of Figure~\ref{figsup:elementos} of this document, are in the equivalent spherical
convention $a-R_E$. The quantities tied to the density and to the simulations, such as the
trajectory of Figure~2 and the injection altitudes of Table~4 of the article, are in mean orbital
geodetic altitude.

The time convention demands the same sharpness, because the work handles three durations that differ
by a few days:

\begin{itemize}
\item The \textbf{archive interval} is the \num{922.7} days (2.526 years) that elapse between the
      first and the last available epoch. It is the quantity against which the errors of the
      dynamical reconstruction are compared.
\item The \textbf{lifetime since deployment} is the \num{926.1} days (2.536 years) elapsed between
      separation of the vehicle, on 15 April 2023 at 06:47 UTC, and re-entry. It is the quantity
      against which the launch companions and FACSAT-1 are compared.
\item The \textbf{simulated orbital lifetime} is counted from the first epoch of the archive until
      the integration crosses the re-entry threshold, and equals 2.53 years in the reference
      scenario. It is the quantity that appears in Tables~2 and~4 of the article, in Table~2S of
      this document and in all the counterfactuals, so that the ratios between scenarios are always
      established between homologous quantities.
\end{itemize}

%%%%%%%%%%%%%%%%%%%%%%%%%%%%%%%%%%%%%%%%%%%%%%%%%%%%%%
\section{Details of the density evaluation}
\label{sup:densidad}

\paragraph{Representative element and sampling.} For each day, the element chosen as representative
is the one occupying the central position in the chronological sequence of that day, understood as
the element of index $\lfloor n/2 \rfloor$ with numbering starting at zero; on days with an even
number of elements the later of the two central ones is taken. The criterion is deterministic and
requires no tie-breaking. From its epoch the orbit is propagated with SGP4 and eight points equally
spaced in time are sampled along one orbital period, from which geodetic latitude, longitude and
altitude above the WGS84 ellipsoid are obtained. NRLMSISE-00 is evaluated at those eight points and
not at a single one, nor only at the mean altitude, which avoids the constant-density approximation:
the thermosphere varies appreciably with altitude, geographic position, local time and above all
with the solar and geomagnetic conditions, and two passes at the same altitude do not imply the same
drag.

\paragraph{Geomagnetic activity vector.}
Geomagnetic activity is not passed to the model as a single daily value but through the
seven-component vector that NRLMSISE-00 admits and that activates its detailed treatment of storms:
the daily $a_p$, that of the current three-hourly interval, those of the three preceding intervals,
and the means of the eight intervals between 12 and 33 hours before and between 36 and 57 hours
before. The indices come from OMNI2 \citep{king2005} and follow the model's standard convention: the
value of the day before the one being evaluated is passed as the daily \Fdiez{} and the centred
81-day mean as $\overline{F}_{10.7}$.

\paragraph{The three day counts.}
The archive spans \num{922.7} days between its first and its last epoch, distributed over 924
calendar days of which fourteen lack published elements entirely. The density is therefore evaluated
on the 910 days for which at least one usable element exists, and the $B_{\mathrm{eff}}$ series is
left with 909 usable days because its computation additionally requires the time derivative of the
semi-major axis, which is lost at the ends of the smoothed series.

\paragraph{Convergence of the orbital sampling.} Eight points per revolution suffice for the daily
average, and the reason deserves explaining, because the density changes a great deal along an
orbit: the relative scatter among the eight samples of a single day has a median of
\SI{32}{\percent}. Decomposed into harmonics of the argument of latitude, however, that variation is
dominated by the first two modes, with median relative amplitudes of \SI{19}{\percent} and
\SI{10}{\percent}, whereas the third and fourth already fall to \SI{1.5}{\percent} and
\SI{1.2}{\percent}. Since the error of a mean over $N$ equally spaced samples is set by the
harmonics aliased at $N$ and its multiples, reducing the sampling to four points, where the error is
determined by the fourth harmonic, shifts the daily mean density by a median of \SI{1.2}{\percent}
and a 95th percentile of \SI{2.8}{\percent}, in agreement with the measured amplitude of that
harmonic. With eight points the responsible term is the eighth, whose amplitude is smaller than that
of the fourth, so that the error of the average stays below \SI{1}{\percent} in all the altitude
bands traversed.

\paragraph{Expressions of the scenario derived from the 2019 forecast.}
The conversion of the smoothed sunspot number into flux uses the empirical relation
\begin{equation}
F_{10.7} = 63.7 + 0.728\,R + 8.9\times10^{-5}R^{2}
\label{eqsup:flujo}
\end{equation}
\citep{hathaway2015,clette2021}, applied to the monthly curve and not to a single value. The
geomagnetic activity of the scenario is set with the climatological relation
\begin{equation}
a_p = 8 + 0.045\,(F_{10.7}-70),
\label{eqsup:ap}
\end{equation}
which reproduces the mean $a_p \approx \SI{12}{\nano\tesla}$ observed at the maximum and falls
towards \SI{8}{\nano\tesla} under minimum conditions. Since the exospheric temperature of the model
is written in terms of $K_p$ whereas the scenarios and the density correction are expressed in
$a_p$, the conversion between the two indices is always carried out with the standard table of
equivalences \citep{matzka2021}, which assigns $K_p = 1.0$ to $a_p = \SI{4}{\nano\tesla}$,
$K_p = 1.7$ to \SI{6}{\nano\tesla}, $K_p = 2.2$ to \SI{8}{\nano\tesla} and $K_p = 2.7$ to
\SI{12}{\nano\tesla}. That conversion is applied both in the counterfactual scenarios and in the
quiet-geomagnetism one.

%%%%%%%%%%%%%%%%%%%%%%%%%%%%%%%%%%%%%%%%%%%%%%%%%%%%%%
\section{Sensitivity to the smoothing scheme}
\label{sup:suavizado}

The choice of the smoothing window with which the orbital series is built is not innocuous, because
the drag response to a storm lasts little more than a day and any window wider than the phenomenon
averages it away entirely. With the daily derivative over the seven-day smoothed series, which is
the procedure adopted in the article, the enhancement of 11 May 2024 reaches $\times3.38$ and the
intense storms give a median of $\times1.92$. Replacing that derivative by a local regression of
$\pm2$, $\pm3.5$ or $\pm7$ days, the first falls to $\times1.78$, $\times1.56$ and $\times1.29$
respectively, and the correlation with $a_p$ drops from $0.25$ to $0.12$.

The practical consequence is that the enhancement amplitudes reported in Section~4.3 of the article
are comparable only with those of works that use an equivalent smoothing. The ordering of the storm
classes by intensity, in contrast, is independent of the window.

The lag of the correlation with \Fdiez{} does not depend on the filtering either: repeating the
analysis with windows of five, seven, nine, eleven and fifteen days, the maximum remains at the same
lag in all five cases and that of the geomagnetic activity remains at zero.

%%%%%%%%%%%%%%%%%%%%%%%%%%%%%%%%%%%%%%%%%%%%%%%%%%%%%%
\section{Statistical treatment and reproducibility}
\label{sup:estadistica}

\paragraph{Block bootstrap of the ballistic coefficient.} The daily $B_{\mathrm{eff}}$ series retains
temporal structure: its autocorrelation decays with a scale of five days: the 909 available values
are equivalent to some 190 independent observations and a resampling that treated them as
independent would underestimate the interval by a factor close to three. Using a moving-block
bootstrap, the half-width of the \SI{95}{\percent} interval goes from \SI{2.2}{\percent} with
independent resampling to \SI{4.5}{\percent} with fifteen-day blocks, to \SI{6.0}{\percent} with
thirty-day blocks and to \SI{7.8}{\percent} with sixty-day blocks, without the median shifting. The
\SI{6}{\percent} corresponding to thirty-day blocks is adopted as the reference figure, that being
the length used throughout the rest of the work.

\paragraph{Null of the correlation bootstrap.}
For the correlations between the drag anomaly and the indices, the blocks of one series are
resampled independently of those of the other, so that the internal autocorrelation of each is
preserved but the association between the two is destroyed under the null hypothesis. Resampling the
pairs jointly would preserve the correlation and would not construct a valid null. Significances are
additionally evaluated with an effective sample size derived from the first-order autocorrelations
of both series, which for the four indices considered is reduced to 632, 582, 540 and 397
independent observations.

\paragraph{Propagation of the systematic uncertainty.} The uncertainty of the ballistic coefficient
separates into two components. The statistical one comes from the bootstrap described above. The
systematic one, much larger, comes from the absolute density level, and to propagate it $10^{5}$
Monte Carlo realisations were carried out with the following model: a lognormal multiplicative
density bias with parameter $\sigma = 0.20$, consistent with the typical error attributed to
NRLMSISE-00 \citep{emmert2015}; a drag coefficient uniform between \num{2.0} and \num{2.4}; and a
mass normal with mean \SI{7.49}{\kilo\gram} and standard deviation \SI{0.05}{\kilo\gram}. It must be
stressed that the \SI{20}{\percent} attributed to the atmospheric model does not correspond to a
well-established probability distribution, so that the intervals derived from it should be read as
central intervals of the adopted uncertainty model and not as confidence levels in the strict sense.

\paragraph{Significance of the periodogram.}
It is estimated with two different references: random permutations, which destroy all temporal
structure, and random-phase surrogates, which preserve the power spectrum and constitute a
considerably more demanding test. Against the former, the false-alarm probability of the 28.3-day
peak is below \num{0.002}; against the latter, the peak ceases to be distinguishable from the
background, with $p = 0.79$.

\paragraph{Randomisation options.}
To facilitate exact reproduction, the options used are stated: \num{5000} resamplings in the
bootstrap of the median ballistic coefficient and in the thirty-day block bootstrap, \num{3000} in
the intervals of the superposed epoch and of the fit of the storm response, 600 permutations and 600
phase surrogates in the spectral analysis, $10^{5}$ realisations in the Monte Carlo propagation,
\num{2000} realisations per altitude in the distribution of orbital lifetimes and \num{2000}
resamplings in the intervals of the storm-merging analysis. All generators are initialised with
fixed seeds declared in the code, and the version of \texttt{pymsis} used is \pymsisver{}, under
Python~\pythonver{}, with the option \texttt{version=0} that selects NRLMSISE-00 rather than the
library default.

%%%%%%%%%%%%%%%%%%%%%%%%%%%%%%%%%%%%%%%%%%%%%%%%%%%%%%
\section{The $B^{*}$ coefficient as an approximation}
\label{sup:bstar}

The orbital elements include the $B^{*}$ coefficient, which is tempting to use as a direct measure of
the ballistic coefficient through $B = \rho_0/(2B^{*})$. The units demand precision, because the
corresponding field of the two-line element does not declare them and $B^{*}$ is frequently
described as dimensionless. In the SGP4 formulation \citep{vallado2006} $B^{*}$ is expressed in
inverse Earth radii and $\rho_0$ is the reference density of the internal atmospheric model, with a
value of \SI{0.157}{\kilo\gram\per\meter\squared} per Earth radius in the canonical units used here,
so that the ratio returns a ballistic coefficient in \si{\kilo\gram\per\meter\squared}.

\begin{figure}[htbp]
\centering
\includegraphics[width=0.98\textwidth]{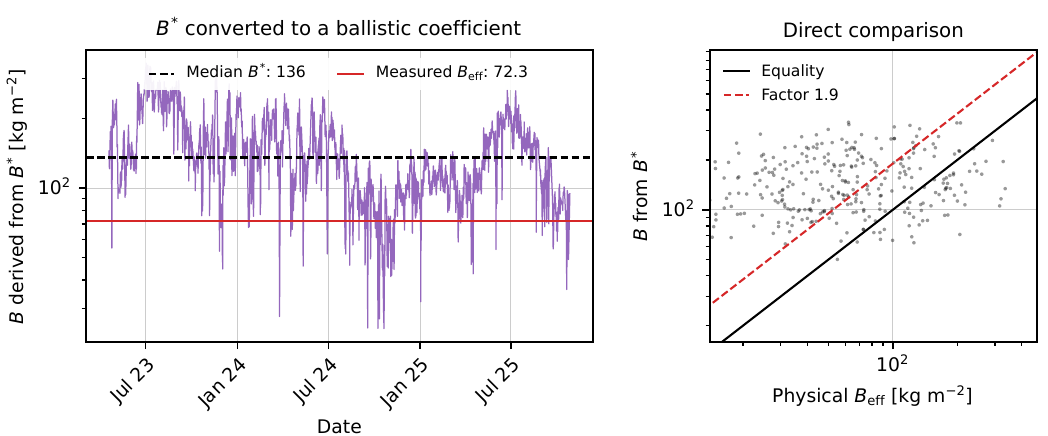}
\caption{Left, ballistic coefficient deduced from $B^{*}$ over the course of the mission, with its
median and with the physical reference value. Right, direct comparison between the two estimates;
the solid line indicates equality and the dashed one the factor of $1.9$ that separates the two
medians.}
\label{figsup:bstar}
\end{figure}

Figure~\ref{figsup:bstar} shows what happens when this is done. The conversion yields a median of
\SI{136}{\kilo\gram\per\meter\squared}, that is, $1.9$ times the physical value obtained by inversion
of the drag, and the correlation between the two estimates is barely $r = 0.24$ on a logarithmic
scale. The reason is well known but frequently forgotten: $B^{*}$ absorbs the deficiencies of the
internal atmospheric model of SGP4 and is not a property of the vehicle. It serves to detect
episodes, such as the peak visible during the Gannon storm, and not to characterise geometries. For
that reason $B^{*}$ is not used in any of the estimates of the main article.

%%%%%%%%%%%%%%%%%%%%%%%%%%%%%%%%%%%%%%%%%%%%%%%%%%%%%%
\section{Evolution of the orbital elements}
\label{sup:elementos}

\begin{figure}[htbp]
\centering
\includegraphics[width=0.94\textwidth]{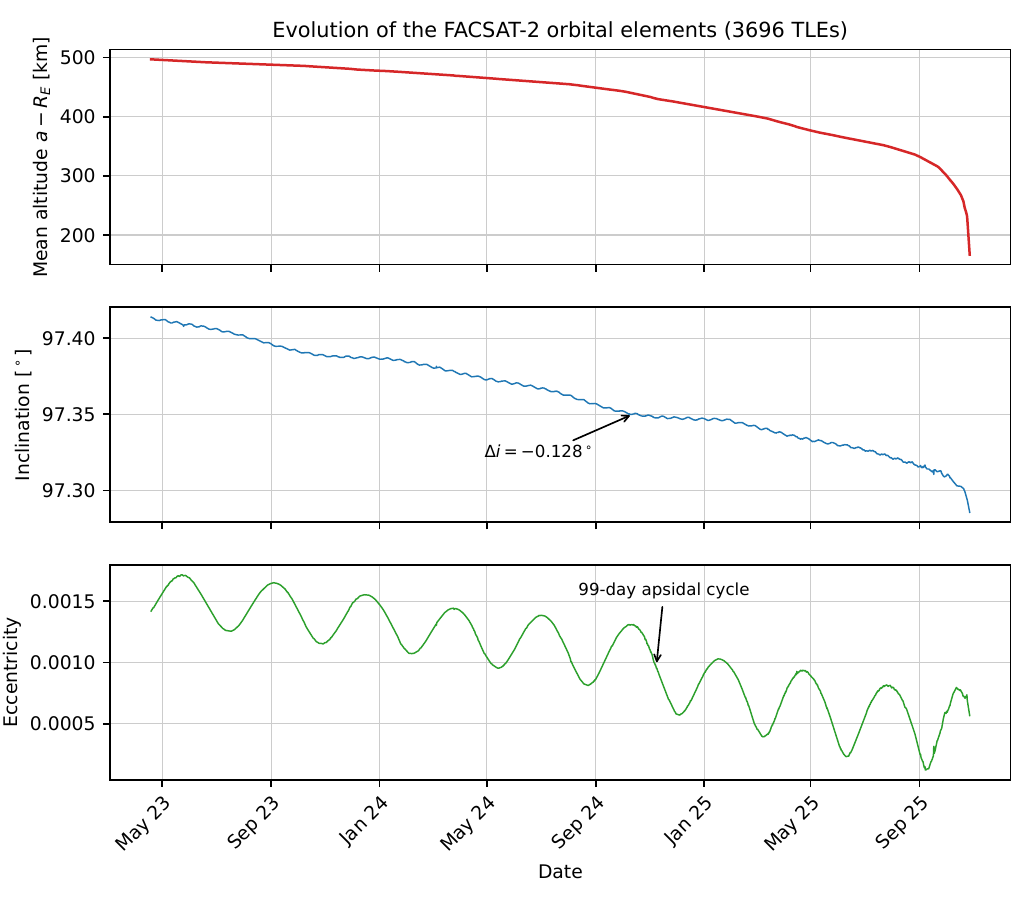}
\caption{Evolution of the mean altitude, the inclination and the eccentricity of FACSAT-2. The
99-day cycle of the eccentricity corresponds to the rotation of the line of apsides induced by the
Earth's oblateness. Altitudes in the equivalent spherical convention $a-R_E$.}
\label{figsup:elementos}
\end{figure}

The archive allows the joint evolution of the altitude, the inclination and the eccentricity to be
followed (Figure~\ref{figsup:elementos}). The three curves tell the same story from different
angles.

The inclination fell from \ang{97.4138} to \ang{97.2857}, that is, \ang{-0.128} over the archive, at
a rate of about \ang{-0.039} per year during the first phase. That drift is real in the elements,
but attributing it to drag without further ado would be hasty, because the rotation of the
atmosphere, gravitational and lunisolar perturbations and the conventions of the SGP4 mean elements
all contribute to it; separating those contributions would require a perturbative propagation that
falls outside the scope of this work.

The eccentricity oscillates with a median period of 99 days while its mean value decays from
$0.00146$ to $0.00057$, that is, the atmosphere circularises the orbit by braking preferentially at
perigee. That period is not arbitrary, since it coincides with the 102 days predicted by the
rotation of the line of apsides due to the Earth's oblateness for this orbit, which constitutes an
internal verification of the coherence of the archive.

None of the three series shows discontinuities, which is the argument used in Section~5 of the
article when ruling out fragmentations or manoeuvres.

%%%%%%%%%%%%%%%%%%%%%%%%%%%%%%%%%%%%%%%%%%%%%%%%%%%%%%
\section{The Gannon superstorm}
\label{sup:gannon}

\begin{figure}[htbp]
\centering
\includegraphics[width=0.84\textwidth]{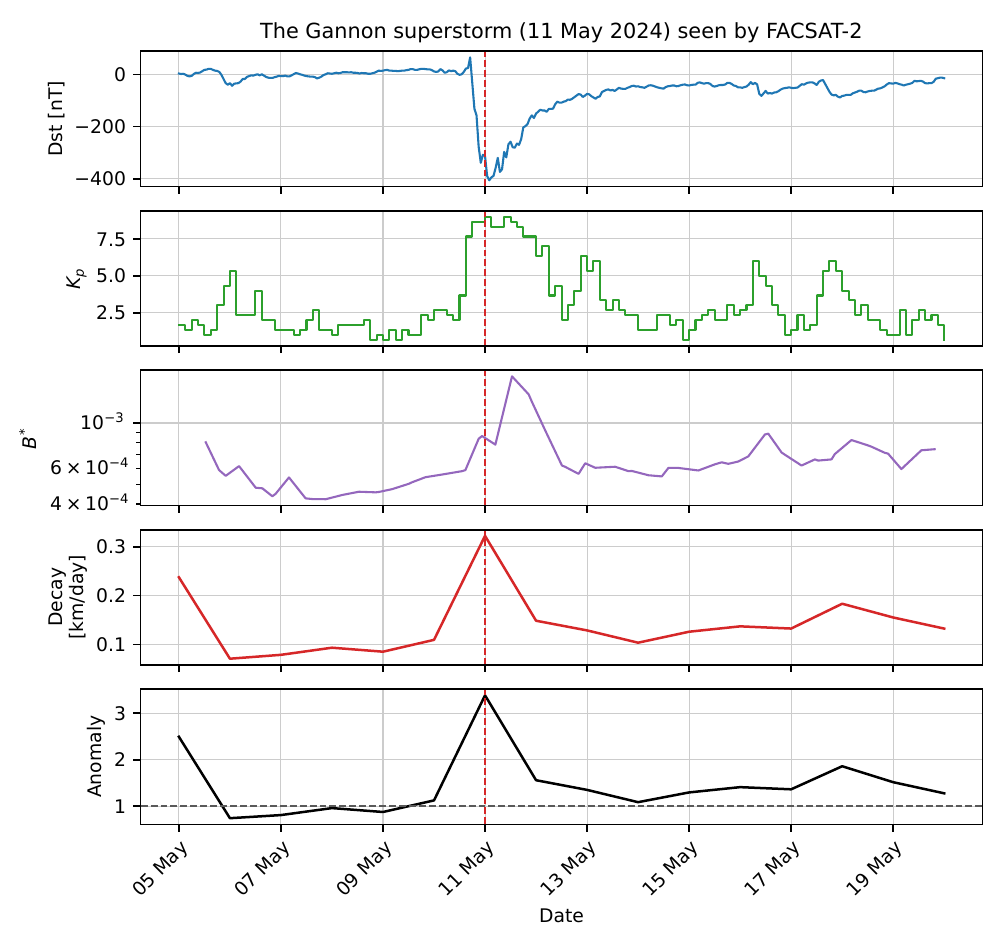}
\caption{The Gannon superstorm seen by FACSAT-2. From top to bottom, hourly Dst index; $K_p$ index
\citep{king2005,matzka2021}, which is a three-hourly index represented here on an hourly grid by
repeating the value within each interval; $B^{*}$ parameter of the orbital elements, expressed in
inverse Earth radii following the SGP4 convention; daily decay rate; and drag anomaly. The
highlighted vertical line marks 11 May 2024.}
\label{figsup:gannon}
\end{figure}

Figure~\ref{figsup:gannon} breaks down the extreme event of the catalogue at hourly resolution. The
sequence is clear: the Dst minimum of \SI{-406}{\nano\tesla} and the $K_p$ maximum of 11 May 2024
coincide with the peak of the $B^{*}$ coefficient of the orbital elements, with a decay rate that
goes from \SI{0.09}{\kilo\meter\per\day} to \SI{0.31}{\kilo\meter\per\day} and with the drag anomaly
reaching $\times3.38$, all on the same day. In this particular event the return to the background
level took somewhat longer than the catalogue average, about five days against the usual two or
three.

That value of $\times3.38$ corresponds to the derivation procedure adopted and should not be read as
a robust physical amplitude: as Section~\ref{sup:suavizado} quantifies, it falls to $\times1.29$
with a $\pm7$-day window.

%%%%%%%%%%%%%%%%%%%%%%%%%%%%%%%%%%%%%%%%%%%%%%%%%%%%%%
\section{Numerical convergence of the integrator}
\label{sup:convergencia}

The drag equation is integrated with second-order Runge--Kutta and a one-day step. That choice
deserves checking, because in the final days the decay exceeds \SI{10}{\kilo\meter\per\day} and the
step ceases to be small compared with the dynamical scale of the problem.

The test was carried out as follows. The density field used in it is not the corrected Jacchia model
of Equation~(4) of the article, but a fit of $\ln\rho$ to a polynomial in altitude with terms in
$\overline{F}_{10.7}$ and $\ln a_p$ built over the same 910 daily orbital mean densities described
in Section~\ref{sup:densidad}, with a residual of \SI{17}{\percent} comparable to that of the former
correction, and with a global factor calibrated to reproduce the observed re-entry date. What the
test measures is therefore the sensitivity of the solution to the numerical scheme, not the accuracy
of the atmospheric model, which is bounded separately in the article.

Repeating the nominal scenario with steps of one, 0.5, 0.25 and 0.1 days, the re-entry date shifts
by 2.5 days on going from one to 0.5, another 0.5 on going to 0.25 and 0.2 more on reaching 0.1: the
successive differences decrease and the solution converges towards some 926 days, counted here from
the first epoch of the archive within this numerical experiment and not to be confused with the
\num{926.1} days elapsed since deployment. The one-day step consequently underestimates the orbital
lifetime by about \textbf{three days}, \SI{0.35}{\percent} of the total lifetime, an error far
smaller than both the 28 days of the out-of-sample validation and that introduced by the density
uncertainty, which moves the simulated lifetime between 2.09 and 3.35 years.

The re-entry threshold turns out to be even less influential. Varying it between \SI{100}{\kilo\meter}
and \SI{150}{\kilo\meter} moves the date by less than a day, as is to be expected from a phase in
which the vehicle traverses that band in barely two days. The solution is therefore converged within
a margin of a few days against both numerical choices.

%%%%%%%%%%%%%%%%%%%%%%%%%%%%%%%%%%%%%%%%%%%%%%%%%%%%%%
\section{Complementary tables}
\label{sup:tablas}

This section gathers three tables that support results discussed in the main article and that have
been moved here for editorial economy. Table~\ref{tabsup:objetos} collects the objects used as
controls and the provenance of their ballistic coefficient; Table~\ref{tabsup:percentiles}
summarises the distribution of orbital lifetime over the ensemble of solar-cycle scenarios; and
Table~\ref{tabsup:modelo} states the parameters of the thermospheric model, of its correction and of
the series treatment, so that the calculation can be reconstructed from the two documents.

\begin{table}[H]
\centering
\caption{Objects used as controls and provenance of their ballistic coefficient. The first three
share injection orbit and deployment date; FACSAT-1 flew five years earlier. All the coefficients
are obtained with the same atmospheric model and three of them by fitting to the re-entry date, so
that the counterfactuals derived from them check the consistency across objects but do not
constitute independent measurements.}
\label{tabsup:objetos}
\small
\begin{tabular}{@{}lccccl@{}}
\toprule
\textbf{Object} & \textbf{Format} & \textbf{Injection} & \textbf{Re-entry} & $B$ [\si{\kilo\gram\per\meter\squared}] & \textbf{Method} \\
\midrule
FACSAT-2 & 6U & \SI{509}{\kilo\meter}, Apr 2023 & 27 Oct 2025 & 72.3 & daily drag inversion \\
TAIFA-1  & 3U & \SI{509}{\kilo\meter}, Apr 2023 & 23 Apr 2025 & 59.5 & fit to the re-entry \\
CIRBE    & 3U & \SI{509}{\kilo\meter}, Apr 2023 & 3 Oct 2024  & 39.3 & fit to the re-entry \\
FACSAT-1 & 3U & \SI{497.8}{\kilo\meter}, Nov 2018 & 3 Jun 2023 & 45.1 & fit to the re-entry \\
\bottomrule
\end{tabular}
\end{table}

\begin{table}[H]
\centering
\caption{Distribution of the orbital lifetime of a vehicle with
$B_{\mathrm{eff}}=\SI{72.3}{\kilo\gram\per\meter\squared}$ over \num{2000} realisations with cycle
amplitude uniform between 115 and 285 sunspots and launch phase uniform within the cycle. The last
column is the fraction of realisations that reach five years, not a historical probability.
Altitudes in the mean orbital geodetic convention.}
\label{tabsup:percentiles}
\small
\begin{tabular}{@{}lrrrr@{}}
\toprule
\textbf{Injection altitude} & \textbf{P5 [yr]} & \textbf{Median [yr]} & \textbf{P95 [yr]} & \textbf{P(life $\ge$ 5 yr)} \\
\midrule
\SI{509}{\kilo\meter} (FACSAT-2) & 1.4 & 4.6 & 8.4 & 0.45 \\
\SI{540}{\kilo\meter} & 2.1 & 6.9 & 11.7 & 0.67 \\
\SI{560}{\kilo\meter} & 2.7 & 8.6 & 17.4 & 0.79 \\
\SI{580}{\kilo\meter} & 3.7 & 10.8 & 22.5 & 0.89 \\
\bottomrule
\end{tabular}
\end{table}

\begin{table}[H]
\centering
\caption{Parameters of the thermospheric model, of its correction and of the series treatment.}
\label{tabsup:modelo}
\small
\begin{tabular}{@{}llr@{}}
\toprule
\textbf{Symbol} & \textbf{Meaning} & \textbf{Value} \\
\midrule
$z_0$      & Reference level of the thermal model & \SI{125}{\kilo\meter} \\
$T_0$      & Temperature at $z_0$ & \SI{355}{\kelvin} \\
$H$        & Effective scale height at \SI{450}{\kilo\meter} & \SI{59}{\kilo\meter} \\
---        & Flux sensitivity (corrected model) & $\times3.15$ per \SI{100}{\sfu} \\
---        & Geomagnetic sensitivity & $\times1.8$ from $a_p=12$ to $a_p=100$ \\
$f_\rho$   & Fitted global density factor & 1.06 \\
$B_{\mathrm{eff}}$ & Measured ballistic coefficient & \SI{72.3}{\kilo\gram\per\meter\squared} \\
$m$        & Nominal mass & \SI{7.49}{\kilo\gram} \\
$C_D$      & Nominal drag coefficient & 2.2 \\
$h_{\mathrm{fin}}$ & Re-entry threshold & \SI{120}{\kilo\meter} \\
$\Delta t$ & Integration step & \SI{1}{\day} \\
$A$        & Amplitude of the 2019 scenario & 115 \\
$t_0$      & Start of the cycle, 2019 scenario & Apr 2020 \\
$b$        & Width of the cycle, $(t_{\max}-t_0)/x_{\max}$ & 4.86 years \\
$x_{\max}$ & Position of the maximum of $g(x)$ & 1.081 \\
$a_p^{\mathrm{quiet}}$ & Level adopted as quiet & \SI{6}{\nano\tesla} \\
---        & Smoothing window of the altitude & \SI{7}{\day} \\
---        & Window of the anomaly baseline & \SI{81}{\day} \\
\bottomrule
\end{tabular}
\end{table}

\clearpage
\renewcommand{\bibname}{References}
%% references_sup.tex -- bibliography of the supplementary material, APA (7th edition)
%% Written by hand for the same reason as references.tex; the keys match.